\documentclass[fleqn,usenatbib]{mnras}

\usepackage{newtxtext,newtxmath}
\usepackage[T1]{fontenc}

\DeclareRobustCommand{\VAN}[3]{#2}
\let\VANthebibliography\thebibliography
\def\thebibliography{\DeclareRobustCommand{\VAN}[3]{##3}\VANthebibliography}

\usepackage{graphicx}	% Including figure files
\usepackage{amsmath}	% Advanced maths commands
\usepackage{subcaption}

\def\mpcoh{{\,h^{-1}\,\rm Mpc}}
\def\hompc{\,h\,\rm Mpc^{-1}}
\def\bigstrut{\vrule width0pt height0.5truecm}
\newcommand{\orcid}[1]{\href{https://orcid.org/#1}{\includegraphics[width=0.7em]{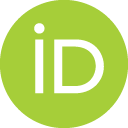}}}
\def\citejap#1{\citeauthor{#1}\ \citeyear{#1}}

\title[Superstructures in the DESI Legacy Survey]{Stacked CMB lensing and ISW signals around superstructures in the DESI Legacy Survey}

\author[Qianjun Hang et al.]
{Qianjun Hang\thanks{E-mail: qhang@roe.ac.uk},
Shadab Alam\orcid{0000-0002-3757-6359},
Yan-Chuan Cai
and John A. Peacock\orcid{0000-0002-1168-8299}
\\
\bigstrut
Institute for Astronomy, University of Edinburgh, Royal Observatory, Blackford Hill, Edinburgh EH9 3HJ, UK\\
}

\pubyear{2021}

\begin{document}
\label{firstpage}
\pagerange{\pageref{firstpage}--\pageref{lastpage}}
\maketitle

% Abstract of the paper
\begin{abstract}
The imprints of large-scale structures on the Cosmic Microwave Background can be studied via the CMB lensing and Integrated Sachs-Wolfe (ISW) signals. In particular, the stacked ISW signal around supervoids has been claimed in several works to be anomalously high. In this study, we find cluster and void superstructures using four tomographic redshift bins with $0<z<0.8$ from the DESI Legacy Survey, and measure the stacked CMB lensing and ISW signals around them. 
To compare our measurements with $\Lambda$CDM model predictions, we construct a mock catalogue with matched galaxy number density and bias, and apply the same photo-$z$ uncertainty as the data. The consistency between the mock and data is verified via the stacked galaxy density profiles around the superstructures and their quantity.
The corresponding lensing convergence and ISW maps are then constructed and compared.
The stacked lensing signal agrees with data well except at the highest redshift bin in density peaks, where the mock prediction is significantly higher, by approximately a factor 1.3. 
The stacked ISW signal is generally consistent with the mock prediction. We do not obtain a significant signal from voids, $A_{\rm ISW}=-0.10\pm0.69$, and the signal from clusters, $A_{\rm ISW}=1.52\pm0.72$, is at best weakly detected. However, these results are strongly inconsistent with previous claims of ISW signals at many times the level of the $\Lambda$CDM prediction. 
We discuss the comparison of our results with past work in this area, and investigate possible explanations for this discrepancy. 
\end{abstract}

% Select between one and six entries from the list of approved keywords.
% Don't make up new ones.

%%%%%%%%%%%%%%%%%%%%%%%%%%%%%%%%%%%%%%%%%%%%%%%%%%

%%%%%%%%%%%%%%%%% BODY OF PAPER %%%%%%%%%%%%%%%%%%

\section{Introduction}

The geodesics of photons in the Cosmic Microwave Background (CMB) are perturbed by their passage through intervening large-scale structures of the universe, generating effects from both spatial and temporal variations in the gravitational potential field, $\Phi({\bf x},t)$. Spatial gradients in $\Phi$ give rise to the gravitational lensing effect, which can be quantified by the lensing convergence $\kappa$:
\begin{equation}
    \kappa(\mathbf{\hat{n}})=\frac{1}{c^2}\int_0^{r_{\rm LS}}\frac{r_{\rm LS}-r}{r_{\rm LS} r} \nabla^2\Phi(\mathbf{\hat{n}},r)\,dr,
\end{equation}
where $r$ is the comoving distance and $r_{\rm LS}$ is the comoving distance to the last scattering surface. This quantity effectively measures the total projected matter density between CMB and today weighted by a distance-dependent kernel for a given angular direction $\mathbf{\hat{n}}$.
The temporal perturbation alters the temperature fluctuations of the CMB, leading to the Integrated Sachs-Wolfe (ISW) effect \citep{1967ApJ...147...73S}:
\begin{equation}
    \frac{\Delta T (\mathbf{\hat{n}})}{T_{\rm CMB}}=-\frac{2}{c^2}\int_0^{t_{\rm LS}} \dot{\Phi}(\mathbf{\hat{n}},t) \, dt,
    \label{eq: ISW}
\end{equation}
where $T_{\rm CMB}=2.725{\rm K}$ is the mean CMB temperature at redshift $z=0$, and $t$ in this expression denotes the look-back time. The ISW effect is intriguing because $\dot{\Phi}\neq0$ in linear theory only in the era of late-time dark energy domination. The measurement of the ISW effect thus provides a dynamical probe for dark energy.
The gravitational potential is related to the matter density fluctuation $\delta$ via the Poisson equation:
\begin{equation}
    \nabla^2\Phi=\frac{3H_0^2\Omega_m}{2a}\delta,
\end{equation}
where $\delta$ is the fractional perturbation in the matter density and $a(t)$ is the dimensionless scale factor.
In practice, the two imprints of potential fluctuations on the CMB can thus be studied using galaxy survey as tracers of the matter density field.

%cross-correlation detection
To measure the gravitational lensing and ISW signals, one approach is the angular cross-correlation between tomographic galaxy density fields and the CMB. The CMB lensing convergence map is reconstructed from the non-Gaussian features of the temperature fluctuations \citep{2000PhRvD..62d3007H, 2003PhRvD..67h3002O,2006PhR...429....1L}. Given the galaxy selection function, the cross-correlation between the CMB lensing convergence and large-scale structure can be detected at $>10\sigma$, both in spherical harmonic space and angular space \citep[e.g.][]{2016MNRAS.456.3213G,2017MNRAS.464.2120S,2018MNRAS.481.1133P,2018MNRAS.480.5386D,2020MNRAS.491...51S,2021MNRAS.500.2250D,2020JCAP...05..047K,hang2021}. These measurements give constraints on the cosmological parameters $\Omega_{\rm m}$ and $\sigma_8$, as well as on the growth rate. Several of these works claim tensions in the $\Omega_m - \sigma_8$ plane compared to the Planck constraints, and similar tensions are also present in cosmic shear measurements \cite[e.g.][]{2020A&A...638L...1J,2021A&A...645A.104A} at the $\sim 3\sigma$ level. 

Similarly, ISW signals have been detected via cross-correlation  \citep[e.g.][]{2003ApJ...597L..89F,2003astro.ph..7335S,2006MNRAS.372L..23C,2008PhRvD..77l3520G,2008PhRvD..78d3519H,2018PhRvD..97f3506S,hang2021}, although the signal to noise is much lower for two main reasons. Firstly, the signal is overwhelmed by the primordial temperature fluctuations in the CMB map; and secondly the signal is concentrated at large scales (low multipoles), where the cosmic variance is largest owing to the small number of independent large-scale modes. One of the most significant detections made using tomographic cross-correlation is by \cite{2018PhRvD..97f3506S}, who reached ${\rm S/N} =4.7\sigma$ by combining several galaxy surveys. The detection of this signal can be used to constrain the dark energy equation of state $w$, e.g. the analysis carried out in \cite{2018PhRvD..97f3506S}.

%super structures literature review of ISW stacking papers and some discussion
\cite{2008ApJ...683L..99G} took an alternative approach of pursuing a stacking analysis that was focused on superstructures. In this work, they averaged CMB temperature maps at the positions of 100 objects identified as voids and clusters that had the most extreme density contrasts as measured using the SDSS LRG sample. By comparison to $\Lambda$CDM simulations, they claimed an excess ISW signal of $4\sigma$ significance.
Subsequently, \cite{2014ApJ...786..110C,2017MNRAS.466.3364C,kovacs2017,2018MNRAS.475.1777K} used stacking techniques and claimed an ISW signal that was higher than the $\Lambda$CDM prediction at moderate significance. \cite{2016ApJ...830L..19N} reported a signal consistent with $\Lambda$CDM using the whole void catalogue, rather than focusing on superstructures.
Most recently, \cite{kovacs2019}, hereafter K19, measured the stacked ISW signal using the DES supervoids with radius $R_v>100\mpcoh$, and found an amplitude relative to the $\Lambda$CDM prediction of $A_{\rm ISW}=5.2\pm 1.6$ in combination with BOSS. In a separate paper, \cite{2021MNRAS.500..464V} measured the stacked CMB lensing convergence signal for the same objects, and found no discrepancy with $\Lambda$CDM. 

The anomalous ISW amplitude from supervoids is of interest in terms of modified gravity, where the screening mechanisms in some theories are less effective in empty regions \citep{2013MNRAS.431..749C}. However, there has not been a satisfactory explanation for this excess signal. \cite{2014ApJ...786..110C} argued that the signal is unlikely to be caused by Sunyaev-Zel'dovich effects, non-Gaussianity, or modified gravity (see also \citejap{2012JCAP...06..042N}). Another possible explanation comes from the AvERA (Average Expansion Rate Approximation) model \citep{2018MNRAS.479.3582B}, which assumes an inhomogeneous expansion rate with $\Omega_{\Lambda}=0$ and predicts a higher overall ISW signal by modifying the growth rate.
However, \cite{hang2021} showed that the AvERA model prediction is inconsistent with galaxy-temperature cross-correlation results, so the ability of the AvERA model to account for the supervoid results is subject to doubt.

One needs to be cautious in interpreting the stacked results. Firstly, the definition of supervoids is not exactly the same in each case. In some cases, voids are defined in 3D density fields based on e.g., the ZOBOV algorithm \citep{2008MNRAS.386.2101N}, whereas in other cases the void definition is based on 2D smoothed density fields \citep[e.g.][]{sanchez2017}. Different void-finding algorithms can lead to different structures being selected. Secondly, the procedures involve various parameter choices such as the initial smoothing scale of the density field and threshold criteria for superstructure selection. If an enhanced ISW signal is to be accepted as a genuine physical effect, it should be robust with respect to these different selection criteria.

Nevertheless, the reported anomalous ISW amplitudes are usually at the 2-3$\sigma$ level, so it remains possible that they are statistical flukes. To clarify the situation, it is useful to use a bigger sample of galaxies for the analysis to beat down the dominant noise from sample variance.
%Aim of this paper
The aim of this work is to repeat the stacking analysis using superstructures in the DESI Legacy Imaging Survey. The large sky coverage reduces the noise due to cosmic variance. We use the galaxy maps produced in \cite{hang2021}, hereafter H21, based on photometric redshifts; the cross-correlation of these maps with the CMB lensing convergence and ISW effect provides a baseline for the ISW amplitude coming from superstructures only. We attempt to adopt the same void finding algorithm as in K19 based on the 2D maps, although  the relatively high thickness of the photometric redshift bins means that our selected superstructures are not exactly comparable to those of K19. In order to reduce confirmation bias, we also adopt a `blind' strategy where we fix our analysis pipeline using mock data based on cosmological N-body simulations, before we run the pipeline on the actual data.

%structure
The paper is organized in the following structure. %Section~\ref{sec: model} describes the linear ISW model from lensing convergence. 
Section~\ref{sec: dataset and simulation} introduces the data used for creating superstructures, the mock galaxy dataset, and the generated lensing convergence and ISW maps. The void finding procedure and covariance matrix is described in Section~\ref{sec: methods}. We compare our superstructure catalogues from the real and mock data in Section~\ref{sec: The void catalogue} and present the stacking results in Section~\ref{sec: results}. Finally, we discuss the results and sum up in Section~\ref{sec: conclusions}.

%\section{Modeling of the stacked lensing and ISW signal}
%\label{sec: model}

\section{Dataset and simulation}
\label{sec: dataset and simulation}
\subsection{DESI Legacy Survey}

We utilize the galaxy density maps in four tomographic bins between $0<z<0.8$ constructed in H21 using the DESI Legacy Imaging Survey. In this section, we briefly describe the procedure by which these maps were constructed.

%DESI Legacy Survey
The DESI Legacy Imaging Survey \citep{DeyLegacy2019} consists of observations from three different projects: divided around ${\rm Dec}=33^\circ$ in J2000 coordinates, the southern hemisphere is observed by the Dark Energy Camera Legacy Survey \citep[DECaLS;][] {2015AJ....150..150F,2005astro.ph.10346T}, wheres the northern hemisphere is covered by Mayall $z$-band legacy Survey \citep[MzLS;][]{2016SPIE.9908E..2CD} and the Beijing-Arizona Sky Survey \citep[BASS;][]{2004SPIE.5492..787W}. The survey covers an area of 17,739 deg$^2$. The data used in this paper come from the publicly available Data Release 8\footnote{\url{http://legacysurvey.org/dr8/}}.

%Galaxy selection
The PSF type objects were excluded from the galaxy selection, and extinction correction was applied to the three optical bands $g$, $r$, and $z$, as well as the WISE \citep{Wright2010} flux $W_1$. The following magnitude cuts were also applied: $g<24$, $r<22$, and $W_1<19.5$, in order to achieve uniform depth over the survey area.
Survey incompleteness was quantified using Bitmasks\footnote{\url{http://legacysurvey.org/dr8/bitmasks}}, with ${\rm bits}=(0, 1, 5, 6, 7, 11, 12, 13)$ masked.
%Photometric redshift calibration
The photometric redshifts were calibrated for selected galaxies using the following spectroscopic samples: GAMA \citep[DR2; ][]{2015MNRAS.452.2087L}, BOSS \citep[DR12; ][]{2015ApJS..219...12A}, eBOSS \citep[DR16; ][]{2020ApJS..249....3A}, VIPERS \citep[DR2; ][]{2018A&A...609A..84S}, and DEEP2 \citep{2013ApJS..208....5N}. In addition, we also included COSMOS \citep{2009ApJ...690.1236I} and DESY1A1 redMaGiC \citep{2018MNRAS.481.2427C} for their highly accurate photometric redshifts. The spectroscopic samples were matched with DECaLS objects based on their nearest neighbours. All samples except the DESY1A1 redMaGiC sample were binned in 3-dimensional grids of $g-r$, $r-z$, and $z-W_1$ with a pixel width of about 0.03. Pixels containing more than 5 objects from the calibration samples were assigned the mean redshift of these objects. The DES samples were then processed in the same way to fill out pixels not assigned with redshift. This 3D grid was used to assign redshifts to $78.6\%$ of the selected Legacy Survey galaxies. The photometric redshifts of the catalogue were compared with that in \cite{Zhou2020}, who applied a random forest method to assign photometric redshifts with a similar set of spectroscopic calibration samples. To improve the accuracy of photometric redshift, we further selected galaxies that have a difference of $|\Delta z|<0.05$ between the two photometric redshifts. This removed a further 23.4\% of the sample.

These galaxies are separated into four tomographic bins: bin 0: $0<z\leq 0.3$; bin 1: $0.3<z\leq 0.45$; bin 2: $0.45<z\leq 0.6$; bin 3: $0.6<z\leq 0.8$. We use {\tt healpix} maps \citep{healpix} with ${\rm nside}=512$ to construct galaxy density maps. The galaxy density is given by $\delta_g=n/\bar{n}-1$, where $n$ is the number of galaxies in each pixel, and $\bar{n}$ is the mean number of galaxies. 

%photometric redshift errors and photo-z distribution
The photometric redshift scatter in each tomographic bin is modelled by a modified Lorentzian of the form
\begin{equation}
    L(x)=\frac{N}{\left(1+((x-x_0)/\sigma)^2/2a\right)^a},
    \label{eq:lorentzian}
\end{equation}
where $N$ is a normalization such that $\int L(x)\, dx=1$; $x_0$ is a shift in the mean redshift; $\sigma$ controls the width of the scatter; and $a$ controls the tail of the scatter. This function is convolved with the raw redshift distribution in order to model the underlying true selection function. Eq.~\ref{eq:lorentzian} provides a good fit to the calibration sample. To account for fainter galaxies, we allow $x_0$ and $a$ to be free parameters, and further impose that the sum of $x_0$ in four redshift bins to be zero. This results in 7 nuisance parameters. In H21 we determined these parameters simultaneously with galaxy bias by fitting the ten galaxy auto- and cross-correlations in spherical harmonic space between the four tomographic bins. We use the best-fit photo-$z$ parameters in this paper, with linear galaxy bias fixed at the minimum $\chi^2$ value. The bias values in the four redshift slices are 1.25, 1.56, 1.53, and 1.83 respectively, assuming a fiducial Planck 2018 cosmology \citep{Planck2018Param}.

%Galactic corrections
There are two main systematic corrections applied to the galaxy density maps. The first one is survey completeness. Pixels with completeness $<0.86$ are masked, and the unmasked pixels are weighted by the inverse of the completeness. Using the ALLWISE total density map as a proxy for stellar density, we also correct systematics near the galactic plane. We introduce a cut in stellar density $N_{\rm star}<1.29\times10^4$\,deg$^{-2}$, and correct for the residual trend of $\delta_g$ with $\log_{10}(N_{\rm star})$ by a 5th-order polynomial in each tomographic slice. We check the cross-correlations between the corrected density maps and the completeness map, as well as the ALLWISE total density map, and confirm that these are consistent with zero.

\subsection{Simulation}
We make use of the MultiDark Planck \citep[MDPL2; ][]{2016MNRAS.457.4340K} simulations with {\em Planck} 2013 Cosmology. The simulation is performed with a $1\,h^{-1} {\rm Gpc}$ box with $3840^3$ particles using the L-Gadget 2 codes. The mass resolution of the simulation is $1.51\times10^9\,h^{-1} M_{\odot}$. The simulation assumes a flat $\Lambda$CDM cosmology with $\Omega_m=0.307$, $\Omega_b=0.048$, $h=0.67$, $n_s=0.96$ and $\sigma_8=0.823$. The dark matter halo catalogue for 32 snapshots between redshift 0 \& 1 is processed using the ROCKSTAR\footnote{\url{https://bitbucket.org/gfcstanford/rockstar}}
phase space halo finder \citep{behroozi13}, in order to construct galaxy lightcones. The simulation is publicly available through the CosmoSim
database\footnote{\url{https://www.cosmosim.org/cms/simulations/mdpl2/}}
\citep{2012MNRAS.423.3018P,2013AN....334..691R}.
Below we describe the procedure used to generate various simulated DESI surveys\footnote{These products are publicly available at \url{https://gitlab.com/qianjunhang/desi-legacy-survey-superstructure-stacking}}. 
 
\subsection{Simulated galaxy light-cones}

\begin{figure}
 \includegraphics[width=0.45\textwidth]{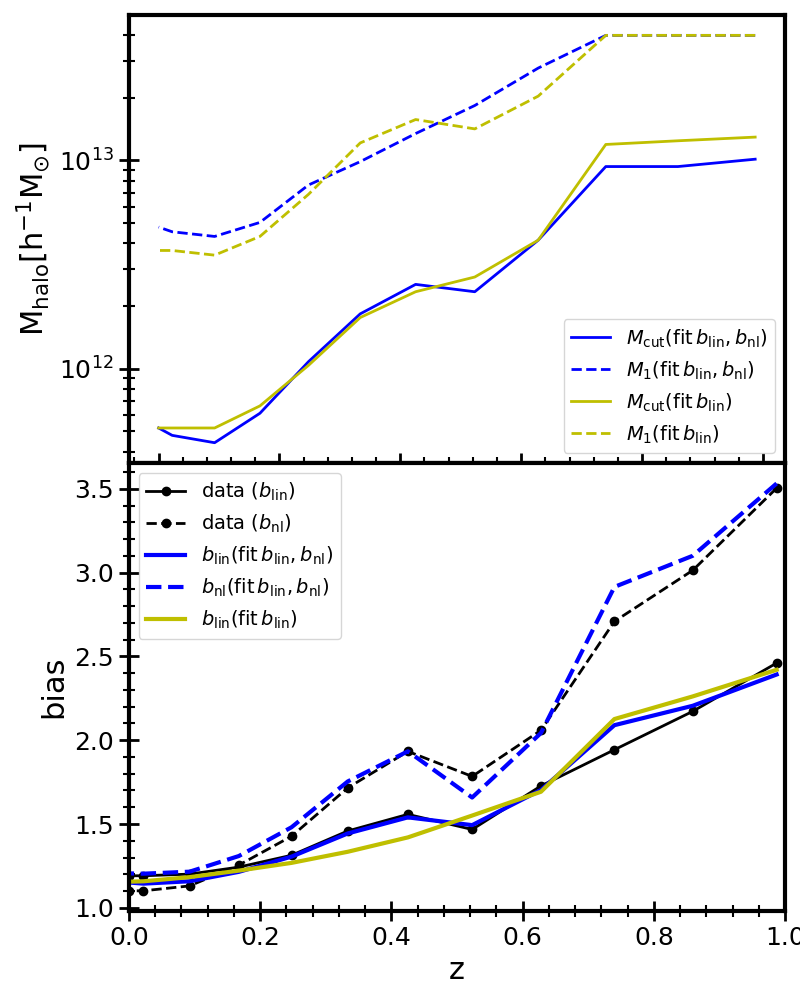}
 \caption{Top panel shows the best fit HOD parameters as the function of redshift used to generate simulated galaxy catalogues. Bottom panel shows the evolution of linear and non-linear bias in mock with coloured lines. The black line shows the best fit linear and non-linear bias obtained for the data from \protect\cite{hang2021}.}
 \label{fig:HOD_z}
\end{figure}

We use the halo occupation distribution (HOD) model to generate simulated galaxy catalogues. We only use the measurements of linear and non-linear bias (H21) to find the best fit HOD parameters. We use a simplified version of the HOD model with only two free parameters corresponding to the characteristic mass of central ($M_{\rm cut}$) and satellite galaxies ($M_1$) as given in following equations:
\begin{center}
    \begin{align}
     p_{\rm cen} &= \frac{1}{2} \mathrm{erfc}\left(\frac{\ln{ M_{\rm cut}-\ln{M_{\rm halo}}}}{\sqrt{2} }\right) \\ 
     \left< N_{\rm sat} \right> &= \frac{M_{\rm halo} - M_{\rm cut}}{M_1},
     \label{eq:HOD}
    \end{align}
\end{center}
where $p_{\rm cen}$ gives the probability of assigning a central galaxy to a halo with mass $M_{\rm halo}$ and $\left< N_{\rm sat} \right>$ gives the mean number of satellite galaxies as the function of halo mass. The actual number of satellite galaxies for any given halo is drawn from a Poisson distribution.
 We use main haloes (i.e. discarding subhaloes) from 32 snapshots between redshift 0 \& 1 and determine the best fit HOD parameters by fitting the 3D galaxy power spectrum with linear and non-linear bias evolution as measured in the data (H21). The linear bias values in our mocks are defined using scales $0.05<k<0.1 \hompc$ and the non-linear bias uses the scales $0.5<k<2 \hompc$. Our best fit parameters are not very sensitive to the limits of scales used to define the linear and non-linear bias. The best fit HOD parameters along with galaxy bias are shown in the Figure~\ref{fig:HOD_z}.
 We have created two sets of mocks, one of which only matches the linear bias, and the other one also has non-linear bias matched. For the scales considered in this project, we confirm that the two mocks do not give rise to significantly different stacking signals from superstructures.

\begin{figure}
 \includegraphics[width=0.47\textwidth]{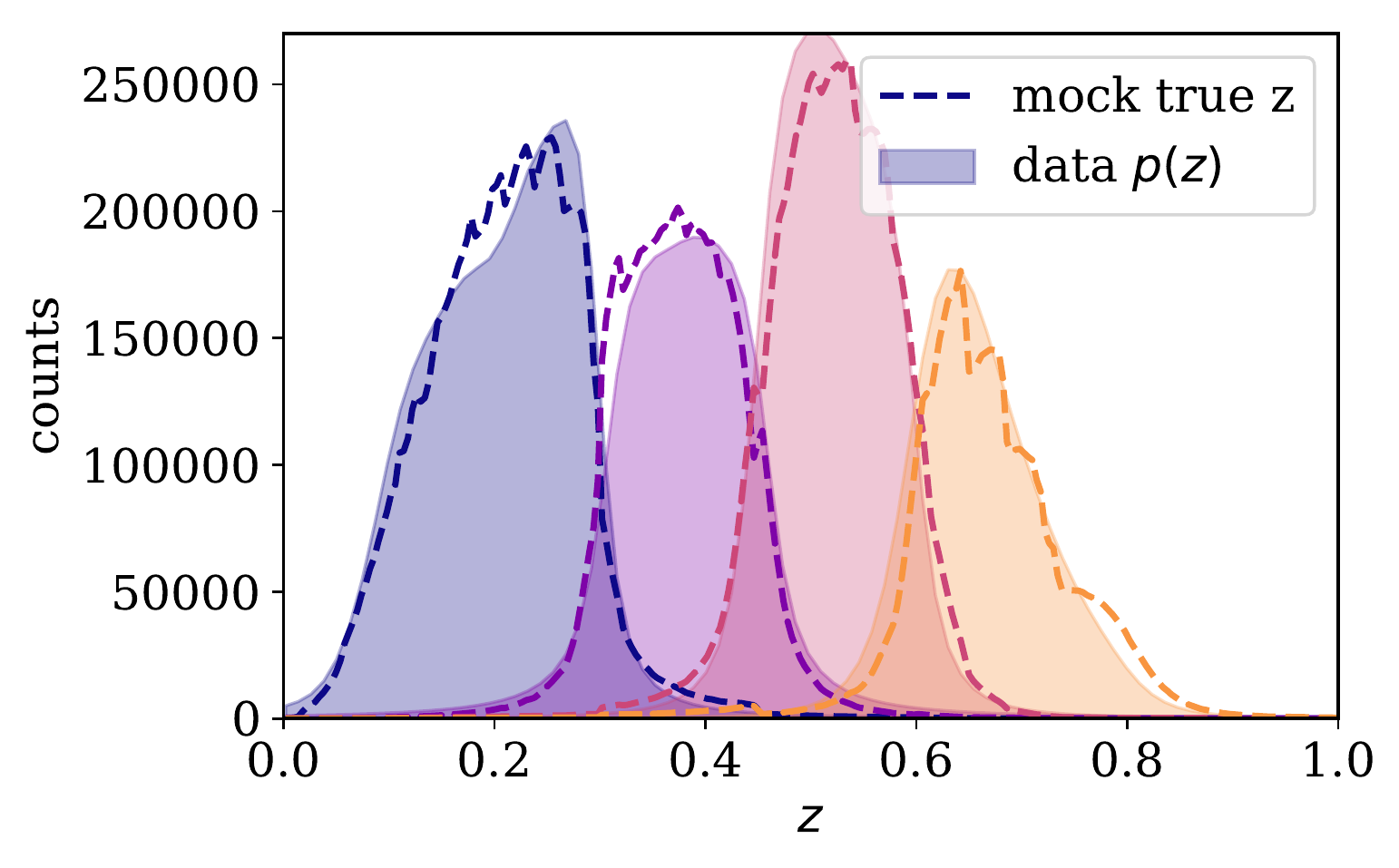}
 \caption{The mock redshift distribution (dashed) is matched to observations by assigning a redshift error  $\delta z$ from the best-fit modified Lorentzian distribution used in \protect\cite{hang2021} and the corresponding best-fit $p(z)$ from data (shaded) by fitting the galaxy auto- and cross-correlations in the four tomographic bins.}
 \label{fig:nz_mock_data}
\end{figure}

%How to match the simulation with data
We then convert our galaxy catalogue into lightcone form by simply repeating the box and placing the observer at the origin in order to extract shells from each snapshot covering the comoving separation between consecutive snapshots. The simulation and data are matched in galaxy number density in each redshift slice. In order to include the photometric redshift effect, we assign to each galaxy a photometric redshift $z_p=z+\delta z$, where $\delta z$ is drawn from the distribution of Eq.~\ref{eq:lorentzian} with the parameters given by the best-fit $p(z)$ in each bin from H21. We then construct our tomographic slices by selecting galaxies in redshift bins using $z_p$. The resulting true redshift distribution is close to the best-fit $p(z)$ from the real data, as shown in Fig.~\ref{fig:nz_mock_data}. The same survey mask is applied to the mock as the DESI Legacy Survey data.

\subsection{Making mock lensing convergence maps}
In order to generate lensing convergence maps that are consistent with our simulated galaxy data, we perform the following integral using the Born approximation:
\begin{equation}
    \kappa(\hat{\theta})=\int_{0}^{r_{\rm max}} \frac{3H_0^2\Omega_m}{2c^2}\frac{(r_{\rm LS}-r)r}{r_{\rm LS}}\delta (r,\hat{\theta})\, dr,
    \label{eq:kappa_born}
\end{equation}
where $r_{\rm LS}$ is the comoving distance between CMB and the lens plane and $r$ is the comoving distance to the lens plane. The  $\delta (r,\hat{\theta})$ is the matter overdensity in the direction $\hat{\theta}$ within a shell of width $dr$ at distance $r$.  To determine $\delta$, we first create particle lightcone using snapshot by repeating the box and extracting a shell of particle at the location of 32 halo snapshot between redshift 0-1. But we have only three particle snapshots ($z \approx0,0.49,1.02$) available compared to 32 halo snapshot. Therefore, for each halo snapshot shell we use the nearest particle snapshot and scale the over-density by ratio of growth at the halo snapshot to the growth at nearest particle snapshot. This gives us $\delta (r,\hat{\theta})$ which is then integrated using equation~\ref{eq:kappa_born}. In principle the full $\kappa$ map should be integrated with $r_{\rm max}=\infty$. But since we are only concerned with the cross-correlation of galaxies with the convergence map, as long as we limit our integral to larger than the maximum galaxy redshift ($z\approx 0.9$) we will obtain unbiased results. Therefore we use $r_{\rm max}$ corresponding to $z_{\rm max}=1.02$ to generate our lensing convergence map. We note that we use a {\tt healpix} pixelization with $n_{\rm side}=512$ to generate our convergence map.

\subsection{Making ISW maps in simulations}
Although the ISW signal arises from the linear evolution of the potential $\Phi$, it has contributions from non-linear evolution. To include both of them, we follow the algorithm presented in \citet{Cai2010, Seljak1996} to compute the time derivative of the potential $\dot\Phi$ in Fourier space using
\begin{equation}\label{eq:dotphi}
\dot{\Phi}(\vec{k},t)=\frac{3}{2}\left(\frac{H_0}{k}\right)^2\Omega_{\rm m}
\left[\frac{\dot{a}}{a^2}\delta(\vec{k},t)+\frac{i\vec{k}\cdot\vec{p}(\vec{k},t)}{a}\right],
\end{equation}
where $a$ is the expansion factor at $z$, $\vec p(\vec k,t)$ is the Fourier transform of the momentum density fluctuation $\vec p(\vec x,t)=[1+\delta(\vec x,t)]\vec v(\vec x,t)$, and $\delta(\vec k, t)$ is the density contrast. We use the full particle data at the three snapshots mentioned above to compute $\dot\Phi(\vec k,t)$ in Fourier space. We then interpolate $\dot\Phi$ in Fourier space according to the linear growth factor $G(a)=D(a)[1-f(a)]$ to obtain $\dot\Phi(\vec k,t)$ at more epochs $t$ between the original snapshots, where the times $t$ are chosen such that their line-of-sight comoving spacing is $100\mpcoh$. The inverse Fourier transform of the above yields $\dot\Phi$ in real space on 3D grids.
Following \citet{Cai2010}, we then use {\sc healpix} to 
tessellate the sky, and follow {\sc healpix} pixel centres along the line of sights to interpolate and integrate $\dot\Phi$ values on grids to obtain the full ISW maps including the non-linear Rees-Sciama effect \citep{Rees1968}. Examples of the power spectra measured from these maps are shown in Fig.~\ref{fig:ISW_map_autoCL_lin_nonlin_compare}. 

\subsection{Quasi-linear ISW maps}
%\YC{(Quasi-linear instead of linear? We are using the non-linear density field to construct the ISW map. The only `linear' assumption is about the coupling between density and velocity. )}

With the expected high signal-to-noise from the galaxy-CMB lensing cross-correlation, we can also use the observed lensing signal around peaks and troughs to predict their corresponding ISW signal directly. This has the benefit of using one observable to predict the other.
Using Eq.~\ref{eq:kappa_born}, we compute the lensing convergence $\kappa$ for each direction $\hat{\theta}$ in each shell between $0<z<1$.
%\begin{equation}
    %\Delta\kappa(\hat{\theta})=\frac{3H_0^2\Omega_m}{2c^2}\frac{(r_{\rm LS}-r)r}{r_{\rm LS}}\delta (r,\hat{\theta})\, \Delta r,
%\end{equation}
The $\kappa$ map can then be converted to the lensing potential, $\psi$, via
\begin{equation}
\kappa_{\ell m}=\frac{1}{2}\ell(\ell+1)\psi_{\ell m}.
\label{eq:kappa to psi}
\end{equation}
This is related to the 3D gravitational potential $\Phi$ via
\begin{equation}
\psi(\hat{\theta})=-\frac{2}{c^2}\int \frac{r_{\rm LS}-r}{r_{\rm LS}r}\Phi(\hat{\theta},r')\,dr'.
\end{equation}
The ISW signal is related to the derivative of the gravitational potential via Eq.~\ref{eq: ISW}.
For the Poisson equation, 
$\nabla^2\Phi=(3/2)H_0^2\Omega_m\delta/a$. In linear theory, it follows that
\smash{$\nabla^2\dot{\Phi}=H(1-f)\nabla^2\Phi$}, or $\dot{\Phi}=H(1-f)\Phi$. Note that $\dot\Phi$ calculated in this way is not fully linear, because the 3D potential includes contributions from non-linearity. The key assumption is that the density and velocity are linearly coupled.

\begin{figure}
\centering
 \includegraphics[width=0.47\textwidth]{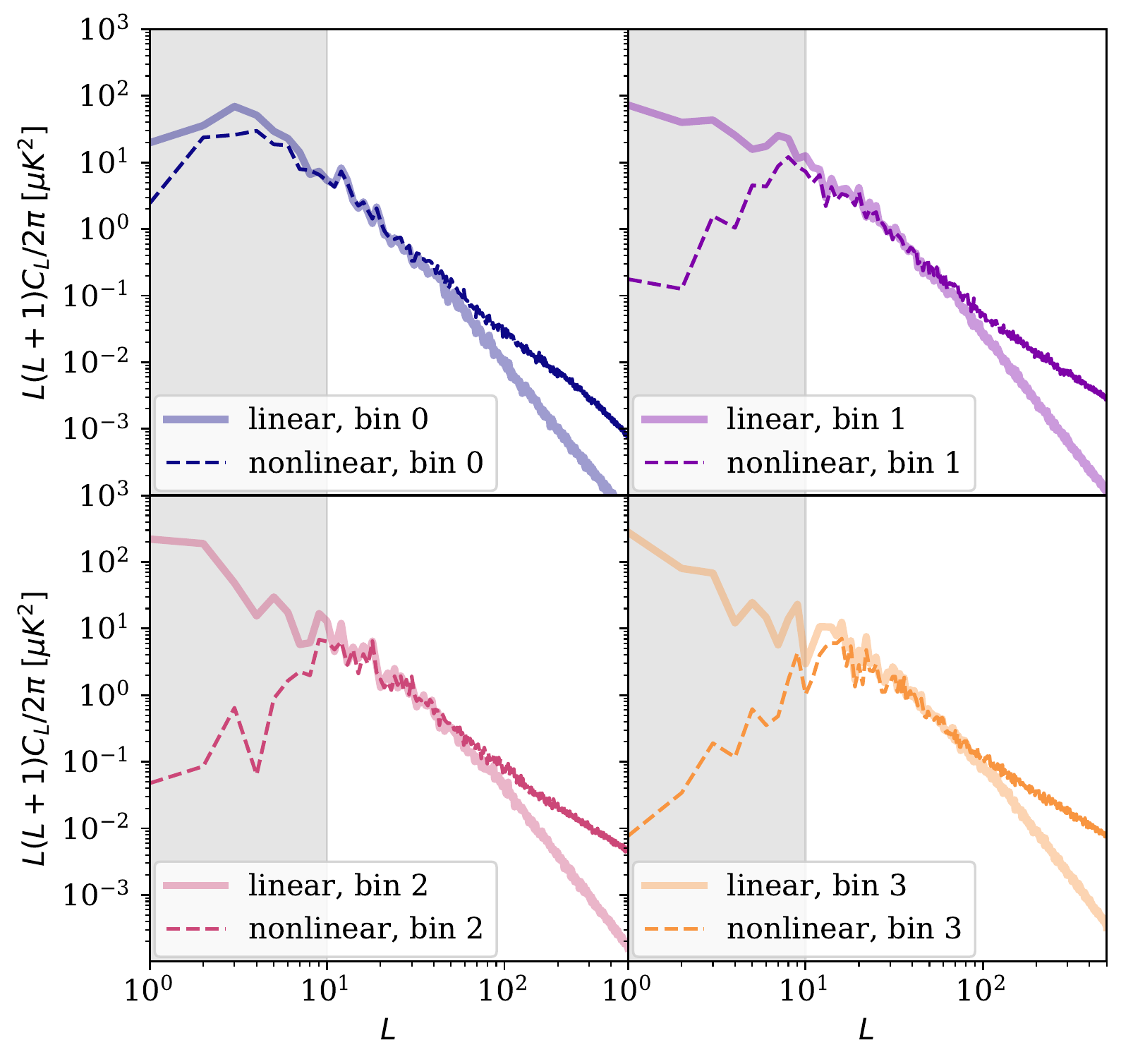}
 \caption{The auto power spectra of simulated ISW maps for the mock catalogue in four tomographic slices. The full non-linear computation is shown with dashed lines and the quasi-linear approximation is shown with solid lines. The grey region indicates the low-$\ell$ range that is removed in the linear map.}
 \label{fig:ISW_map_autoCL_lin_nonlin_compare}
\end{figure}

Given a thin shell centred around redshift $z_0$ with edges $[z_0-\Delta z, z_0+\Delta z]$, one can make the approximations
\begin{align}
&\psi(\hat{\theta},z_0)\approx-\frac{2}{c^2}\frac{r_{\rm LS}-r_0}{r_{\rm LS}r_0}\frac{c}{H(z_0)}\int_{z_0-\Delta z}^{z_0+\Delta z}\Phi(\hat{\theta},z)\,dz,\\
&\Delta T(\hat{\theta},z_0)\approx-T_0\frac{2}{c^2}a(z_0)\left[1-f(z_0)\right]\int_{z_0-\Delta z}^{z_0+\Delta z}\Phi(\hat{\theta},z)\,dz.
\end{align}
Combining these two equations we have
\begin{equation}
\Delta T(\hat{\theta},z_0)\approx T_0 a(z_0)\left[1-f(z_0)\right] \frac{r_{\rm LS}r_0}{r_{\rm LS}-r_0}\frac{H(z_0)}{c}\psi(\hat{\theta}).
\label{eq:delta T linear}
\end{equation}
We obtain the ISW map for each of the 30 shells, where $r_0$ is the comoving distance to the shell centre. These maps are then added together to produce the final (noise-free) ISW map.  
The comparison of the power spectra of the quasi-linear and full ISW maps is shown in Fig.~\ref{fig:ISW_map_autoCL_lin_nonlin_compare}. We can see that the two maps are most consistent in the range of $10<\ell<40$. At scales $\ell\lesssim10$, the linear map gives unphysical modes whose amplitudes are much larger than the full computation. At smaller scales, where $\ell>40$, the full computation gives a higher amplitude than the quasi-linear case. In the stacking analysis, we are mostly interested in structures of a few degrees, corresponding to $\ell\sim 100$.

%\YC{(We may not need the following paragraph and Appendix A for the paper, as they are not relevant? But they are useful for the thesis.)}

%An alternative method based on non-linear density evolution is presented in Appendix~\ref{apdx: method2}. This method assumes spherical symmetry of the stacked void profile. This is generally the case for 3D voids. Void finders based on 2D density slices preferentially find elongated voids \citep{kovacs2017}. For our case, the redshift slice is $400-800\mpcoh$ in width, the voids found are more likely to be `tunnels', and spherical symmetry is not applicable. Therefore, we use our simulations to predict the stacked ISW signal, which captures the non-linearity of the density evolution. As a comparison, we use this method to compute the linear theory prediction of the same signal, and demonstrate their differences.

%The plot for Cl comparison

\section{Methods}
\label{sec: methods}

\subsection{Blind analysis}
Given the low significance of the tentative ISW signal from stacking, we wish to conduct the calibration for the selection of superstructures strictly prior to the unveiling of the stacked temperature results, i.e. a blind analysis. For this, we separate our analysis into two steps: (1) we finalise our selection for superstructures based on the galaxy number density maps and their cross-correlations with the CMB lensing convergence map. The stacking with CMB lensing map is expected to yield much higher signal-to-noise, thus it provides a benchmark for the calibration of our superstructures. (2) We then `unblind' by applying the same stacking with the same catalogues of superstructures for the CMB temperature map.

\subsection{Void finder}

We follow the void finder algorithm in \cite{sanchez2017}. The finder takes the following steps:

\begin{enumerate}
    \item Given the 2D density fluctuation on \texttt{healpix} maps with $\delta=n/\bar{n}-1$, we first apply a Gaussian smoothing of $\sigma=20\mpcoh/d(z)$, where $d(z)$ is the mean comoving distance to the tomographic slice. We then define pixels with $\delta<\delta_*$ as potential void centres. In practice, we fix $\delta_*$ to pick out the lowest $10\%$ of the smoothed pixels, which is around $\delta_*=-0.2$.
    \item Starting from the lowest density pixel in the potential void centres, we compute the mean density $\delta_i$ inside circular shells of radii $R_i$ and $R_i+\Delta R$ for each $R_i\in \{R\}$. $\Delta R$ is chosen to be $1\mpcoh$. Once $\bar{\delta}_i>0$ is encountered for the first time, we register $R_v=R_i+\Delta R/2$ as the void radius. In practice, we use the \texttt{healpy.query\_disc} function to find pixels within a disc of angular size $\theta_i=R_i/d(z)$.
    \item Once the void is found, we check the potential void centre list, and exclude any centre that is inside the existing void. 
    \item We then update the list of potential void centres and repeat steps (ii)-(iii) until the list is exhausted.
\end{enumerate}

\begin{figure*}
    \centering
    \begin{subfigure}[b]{0.49\textwidth}
    \includegraphics[width=\textwidth]{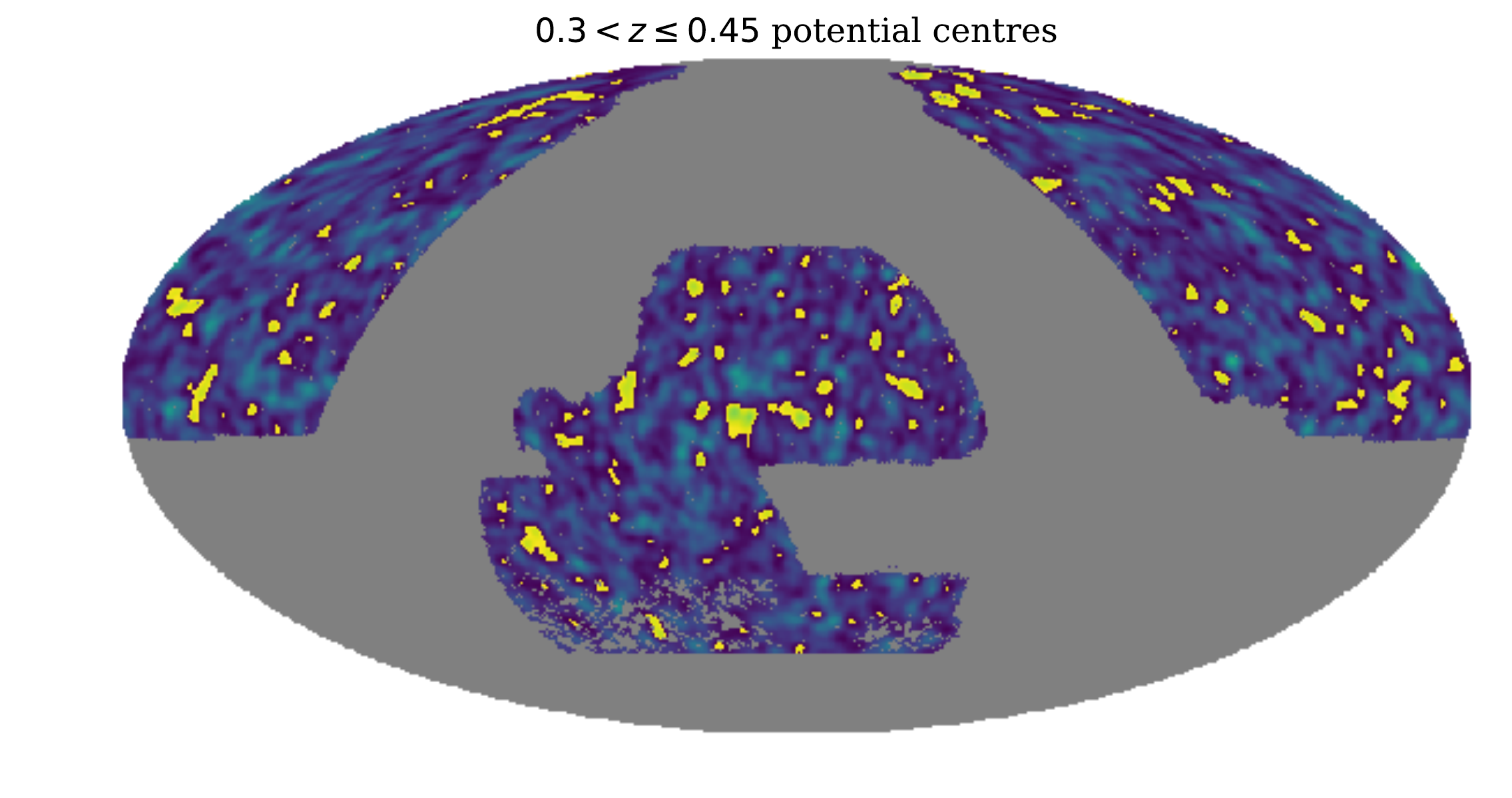}
    \end{subfigure}
    ~
    \begin{subfigure}[b]{0.49\textwidth}
    \includegraphics[width=\textwidth]{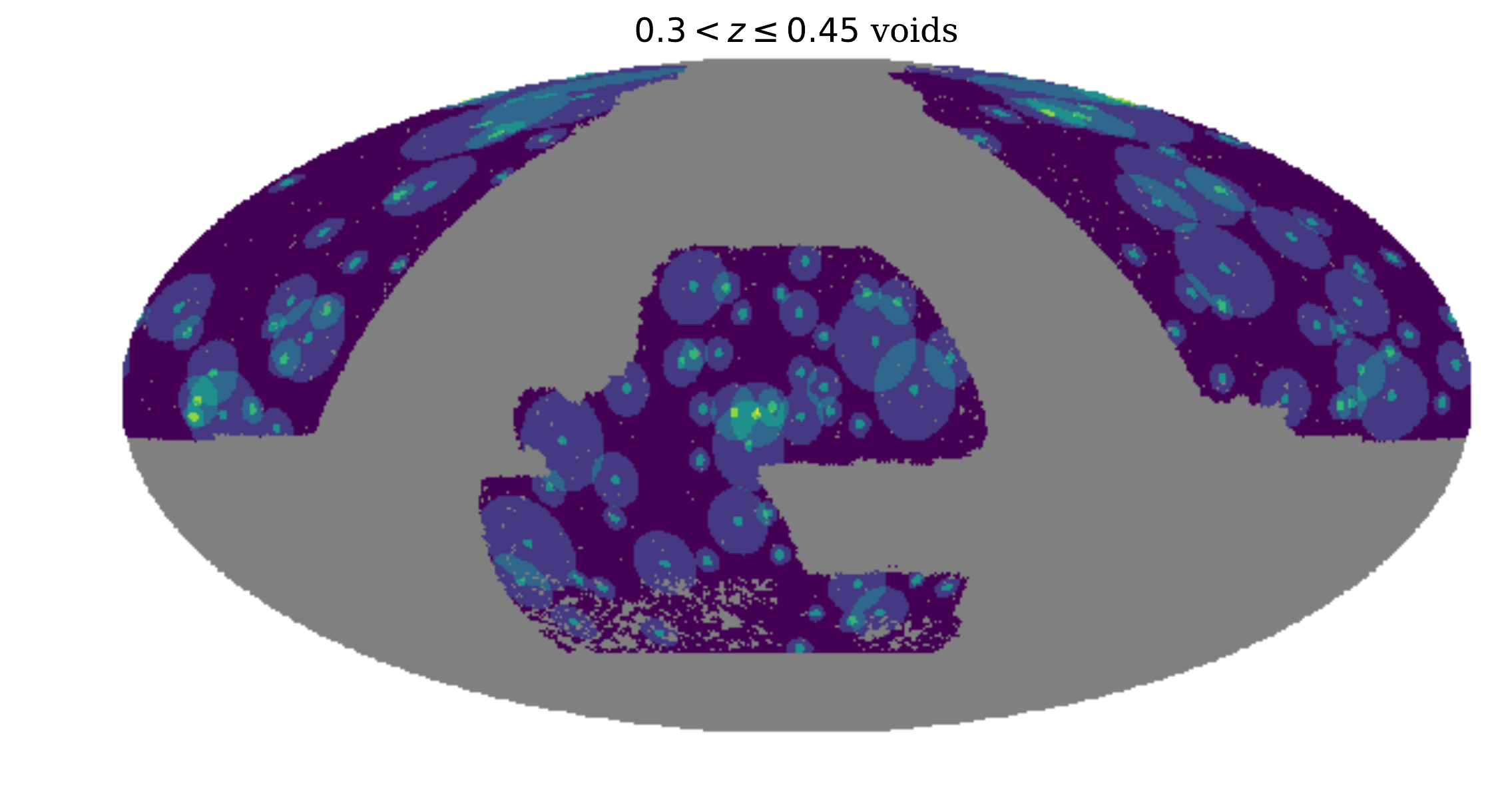}
    \end{subfigure}
    \caption{An example of the void finding procedure using the tomographic slice in redshift range $0.3<z\leq0.45$. Left: the highlighted pixels correspond to the potential void centres, selected on the smoothed density map with $\delta<\delta_*$. In this case, $\delta_*=-0.14$. The background intensity map shows the density fluctuation in this slice. Right: The resulting void centres (shown highlighted dots) and their void radius (shown in fainter circles). Notice that some voids at survey boundaries are excluded. Also notice that voids can overlap, in cases where deeper voids can be found inside shallower voids.}
    \label{fig:void finding procedure}
\end{figure*}

The free parameters in this finder algorithm are the initial smoothing $\sigma$ and the density cut $\delta_*$. A larger $\sigma$ will result in the merging of smaller voids, and could lead to higher signal to noise \citep{sanchez2017,kovacs2019}. 
As a result of merging voids and the hierarchical void-finding procedure, the void catalogue can be different. Increasing $\delta_*$ would include shallower voids. However, this should not affect any deeper voids found with a lower $\delta_*$. It is possible to find small but deep voids embedded in large shallower voids. We choose $\{R\}$ in the range $1\mpcoh\leq R\leq 300\mpcoh$, with an increment of $2\mpcoh$ between each step. After we obtain the void sample, we further exclude voids that have less than $70\%$ of their area inside the survey mask. An illustration of the procedure is shown in Fig.~\ref{fig:void finding procedure}.

A major difference between this work and \cite{kovacs2019} is that our redshift slice is much thicker whereas they used slices of comoving size $100\mpcoh$.
In \cite{kovacs2019}, due to the thin redshift slice, they also pruned overlapping voids between different redshift bins by shifting the bin edges a few times. Thus, although the void finding algorithm is defined in 2D, their void catalogue is comparable to those found using 3D algorithms. We do not carry out this procedure here because we expect that the structures in the four tomographic bins are dominated at distinctive redshifts and thus not strongly correlated. The voids found here are likely to be `tunnels' rather than spherical objects.

To find clusters, we apply an identical procedure to an inverted density map. Due to the lognormal shape of the smoothed density distribution of each map, we select the densest 5\%, instead of 10\%, pixels as potential centres. This choice gives similar numbers of clusters and voids in the final sample.

%Sky position

In order to obtain the stacked signal at the position of these superstructures, we rotate the map (in this case, the map can be galaxy density, lensing convergence, or temperature fluctuation) at the pixel level to place each superstructure centre at $(\theta,\phi)=(0,0)$. We then stack the rotated maps scaled by the void radius $R_v$, on a grid with $0\leq R \leq 3R_v$.
To account for masks, we also perform the same rotation to the mask for each void. The stacked map is obtained by 
\begin{equation}
    P^{\rm stack}=\frac{\sum_i P^{\rm map}_i}{\sum_i P^{\rm mask}_i},
\end{equation}
where $P^{\rm map}_i$ is the pixel value for the map for $i$-th void, and $P^{\rm mask}_i$ is that for the mask. We extract the isotropic radial profile for these stacked images. Given the angular bins $\{\theta\}$, we measure the average signal in the ring between radii $\theta_i$ and $\theta_{i+1}$, and assign the value to the middle of the angular bin. 

\subsection{Covariance matrix}

We use three methods to estimate the covariance matrix for the stacked signal to account for the noise on the background CMB map as well as the foreground superstructure positions.

To capture the CMB noise, we generate 1000 random CMB maps with ${\rm nside}=512$ using the measured pseudo CMB temperature auto power spectrum, corrected by the fraction of sky lost due to the mask $\hat{C}_{\ell}=C_{\ell}/f_{\rm sky}$. The maps are then generated using the \texttt{synfast} function in the \texttt{healpy} package applied to $C_{\ell}$, and multiplied by the {\em Planck} 2018 CMB mask. For comparison, we also use the {\em Planck} best-fit $\Lambda$CDM CMB power spectrum\footnote{\url{http://pla.esac.esa.int/pla/\#cosmology}}, accounting for the pixel window function and the ${\rm FWHM}= 5\,{\rm arcmin}$ circular Gaussian smoothing. These methods give a consistent covariance matrix. We repeat the same stacking process for superstructures in each redshift slice on each of the random CMB maps and extract the averaged radial profile. The covariance matrix is computed by
\begin{equation}
C_{ij}=\frac{1}{N}\sum_{s}^N(x^s_i-\bar{x}_i)(x^s_j-\bar{x}_j),
\label{eq:covariance}
\end{equation}
where $N=1000$ is the sample size, $x^s_i$ is the measurement of $i$-th data component in the $s$-th sample, and $\bar{x}_i$ is the mean measurement of the $i$-th component.
The inverse covariance is corrected by the Hartlap factor \citep{2007A&A...464..399H} with $C^{-1}_{ij}=(N-p-1)/(N-1)\langle C^{-1}_{ij}\rangle$, where $p=15$ is the length of the data vector.

To estimate the errors due to the fluctuations of the foreground galaxy sample, we generate 1000 sets of random void (cluster) positions for each redshift bin within the survey mask, and compute the stacked signal on the {\em Planck} 2018 CMB temperature map \citep{PlanckT2018}. It should be noticed that this assumes no correlation of the positions of the voids (clusters), which is in general not true: there will be close pairs of clusters, while it is unlikely to find two voids that are close to each other. Nevertheless, this method provides a rough estimate of the foreground random error. The covariance is computed using Eq.~\ref{eq:covariance} and the inverse covariance is corrected by the Hartlap factor.

Finally, we estimate the covariance matrix from Jackknife subsampling by excluding one void (cluster) at a time in the given redshift bin. The sample size is equal to the number of voids (clusters) in each bin, $N_J$. The resultant covariance matrix from Eq.~\ref{eq:covariance} is multiplied by $(N_J-1)$ to account for the correlation between different Jackknife samples. %The smallest Jackknife sample size is in the lowest redshift bin, where $N_J=41$ for data and $N_J=39$ for the simulation. 
The Jackknife covariance matrix is noisy with small sample size, i.e., in the lower redshift bins.

\begin{figure}
\centering
 \includegraphics[width=0.45\textwidth]{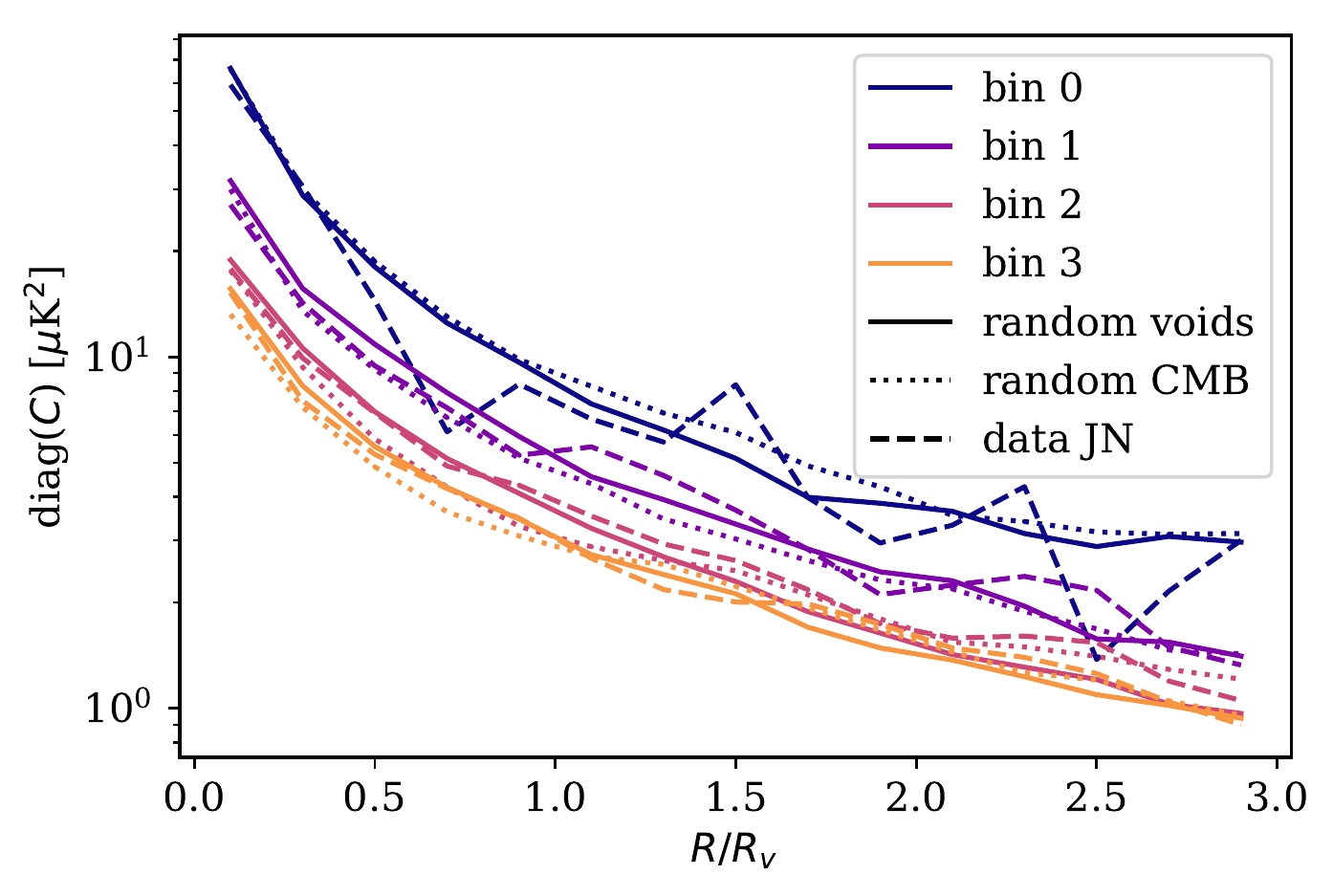}
 \caption{The diagonal elements of the covariance matrices (in $[\mu {\rm K}^2]$) for the radial ISW stacked profile in each redshift bin (shown in different colours). The dotted lines show that from 1000 random CMB samples using the void positions in data, the solid lines show that from 1000 sets of void positions using the real CMB map, and the dashed lines show the Jackknife error from the actual data. }
 \label{fig:random_CMB_voids_covariance_diag}
\end{figure}

The comparison of the diagonal elements of the three covariance matrices for the void sample is shown in Fig.~\ref{fig:random_CMB_voids_covariance_diag}. For the cluster sample, the covariance is similar but with different number of objects in each bin. In all cases, there is close agreement between the three methods. Due to the small Jackknife sample size in bin 0, the diagonal elements are noisy compared to the other two methods. From here on, we will use the covariance matrix estimated from random superstructure positions in our following analysis. The Jackknife covariance is used for the case of the stacking of all superstructures. 

\section{Superstructures}
\label{sec: The void catalogue}

\begin{figure}
 \includegraphics[width=0.47\textwidth]{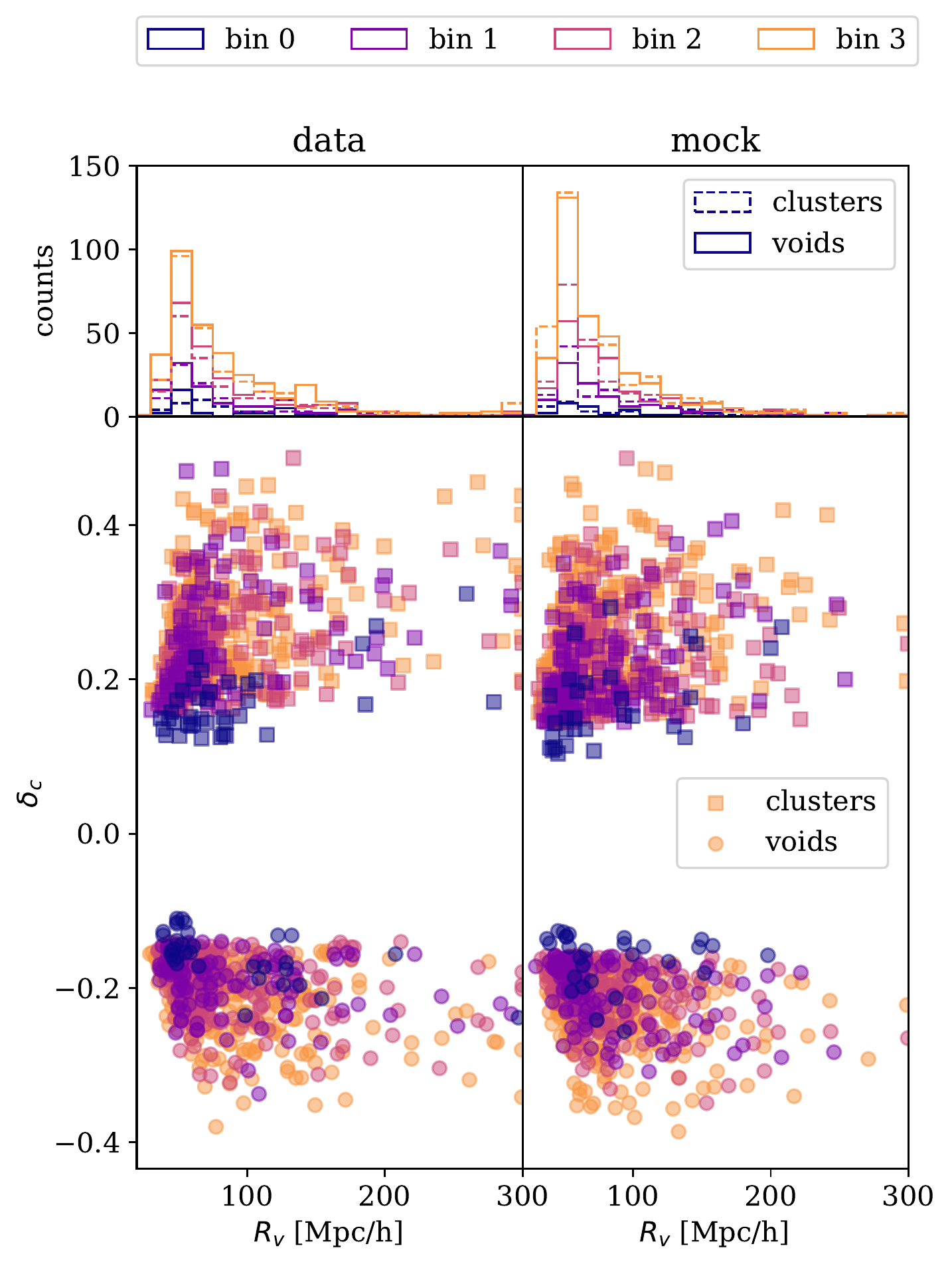}
 \caption{Superstructure size ($R_v$) and central density ($\delta_c$) in data and the mock for each redshift bin. The radius is defined as when the mean density measured within a ring of central radius $R$ and width $1\mpcoh$ first become positive.}
 \label{fig:void radius}
\end{figure}

A summary of the numbers of voids and clusters found in each redshift bin is shown in Table~\ref{tab:void number} for both the mock and real data. 
In general, the data and the mock show good consistency in terms of the number of voids found and in the distribution of void radius. For clusters, the density cut $\delta_*$ in the finder algorithm is slightly larger in data compared to mock, and the number of clusters found is smaller especially in bin 2 and bin 3.
The distribution of the radius in comoving length and central density (in the smoothed map) of these superstructures is shown in Fig.~\ref{fig:void radius}. The majority of the superstructures found have a radius of around $60\mpcoh$, with an extended tail towards $R_v\sim300\mpcoh$. There is a small number of clusters in data that saturate at the maximum radius. It is pointed out in \cite{kovacs2017} that there is an anti-correlation between the depth and the size of the superstructures. There is, however, no clear trend in the voids and clusters found here. The minimum $R_v$ at fixed central density increases with the central density becoming more extreme.

The stacked galaxy density profiles are shown in the upper panel of Fig.~\ref{fig:stacked profile kappa} for both voids and clusters. The agreement between mock (solid bands) and real data (circles for voids and squares for clusters) is good. The dotted lines show the profile divided by linear galaxy bias in each case. The agreement between data and simulation using linear bias is expected for voids, as discussed in \cite{2017MNRAS.469..787P}.
At $R>R_v$, the stacked density profile changes sign and peaks at $R\sim1.3R_v$, before falling to zero at larger scales. This suggests that on average, the voids found are surrounded by overdensities and clusters are surrounded by underdensities.

%table of number of voids
\begin{table*}
 \caption{Summary of various parameters used in superstructure finding and the number of superstructures in each redshift slice for the mock and data respectively. The first two rows show the mean redshift computed from the best-fit redshift distribution and the linear galaxy bias in H21. The third row shows the smoothing scales for the density maps in units of degrees, which correspond to a comoving length of $20\mpcoh$ for each slice. The last few rows show the density cut, where $\delta<\delta_*$ ($\delta>\delta_*$) are selected as potential void (cluster) centres, as well as the number of objects found in each bin, after excluding those that have less than $70\%$ of their area inside the survey mask.}
 \label{tab:example}
 \begin{tabular}{lcccc}
  \hline
  Redshift bin & $0< z\leq 0.3$ & $0.3< z\leq 0.45$ & $0.45< z\leq 0.6$ & $0.6< z\leq0.8$\\
  \hline
  \textbf{Mock}&&&&\\
  Mean redshift & 0.210 & 0.376 & 0.521 &  0.667\\
  linear bias & 1.19 & 1.40 & 1.49 & 1.76\\
  Smoothing scale [deg] & 1.92 & 1.12 & 0.84 & 0.68\\
  $\delta_*$ (voids) & $-$0.11 & $-$0.14 & $-$0.14 & $-$0.15\\
  $N$ (voids) & 28 & 108 & 209 & 364\\
  $\delta_*$ (clusters) & 0.11 & 0.15 & 0.15 & 0.16\\
  $N$ (clusters) & 31 & 119 & 230 & 378\\
  \hline
  \textbf{Data}&&&&\\
  Mean redshift & 0.207 & 0.376 & 0.522 & 0.663\\
  linear bias & 1.25 & 1.56 & 1.52 & 1.83\\
  Smoothing scale [deg] & 1.93 & 1.09 & 0.84 & 0.69\\
  $\delta_*$ (voids) & $-$0.11 & $-$0.14 & $-$0.14 & $-$0.15\\
  $N$ (voids) & 33 & 111 & 223 & 332\\
  $\delta_*$ (clusters) & 0.12 & 0.16 & 0.16 & 0.18\\
  $N$ (clusters) & 38 & 97 & 185 & 282\\
  \hline
 \end{tabular}
 \label{tab:void number}
\end{table*}

%plot of void radii in four bins

\section{Results}
\label{sec: results}

\subsection{Stacked lensing map}

We stack the {\em Planck} 2018 lensing convergence map \citep{Plancklens2018} with $\ell_{\rm max}=2048$ and the simulated lensing convergence map at superstructure positions in real and mock data respectively. Prior to stacking, we smoothed the lensing maps with a Gaussian kernel with ${\rm FWHM}=1^{\circ}$ to suppress the small scale power for the purpose of map rotation at the pixel level, and this is done consistently in both data and simulation.

\begin{figure*}
\centering
 \includegraphics[width=\textwidth]{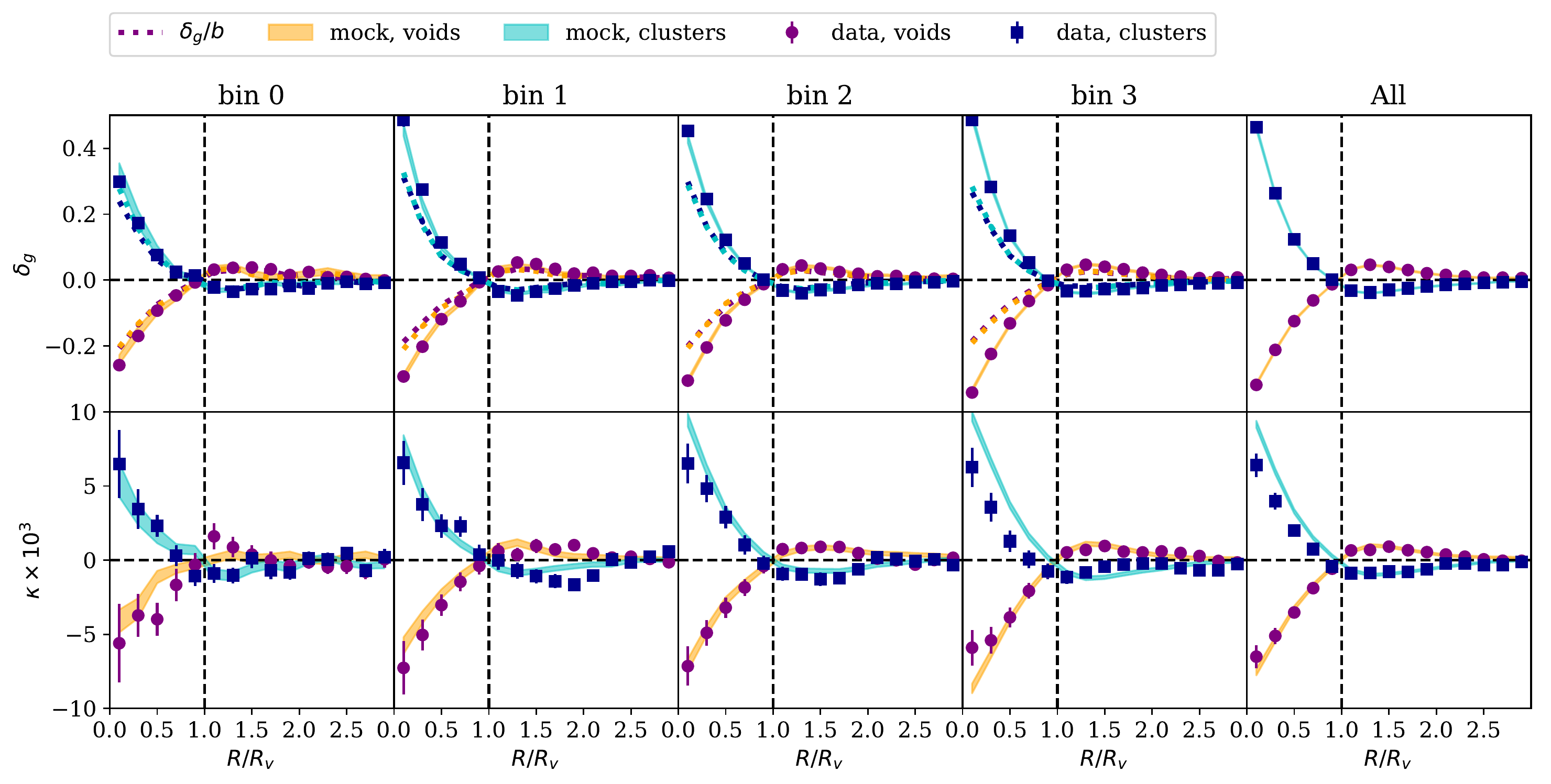}
 \caption{Upper panel: The averaged radial profile of stacked galaxy density in each redshift bin at the superstructure positions found in the mock (solid band) and data (points). The dotted lines show the mean profile divided by the linear galaxy bias measured in the mock and data respectively. Lower panel: The averaged radial profile of stacked CMB lensing convergence. The lensing map have been smoothed by a Gaussian kernel of ${\rm FWHM}=1^{\circ}$ to suppress the small scale power for the purpose of map rotation at the pixel level. The error bars come from Jackknife sampling of the superstructures in each redshift bin.}
 \label{fig:stacked profile kappa}
\end{figure*}

The lower panel of Fig.~\ref{fig:stacked profile kappa} shows the stacked radial profile of the $\kappa$-map. 
Similar to the case of the stacked galaxy density profile, the change of sign with a peak at $R\approx1.3R_v$ is also present in the stacked $\kappa$ profiles. 
For voids, the real and mock datasets show good consistency in general. For clusters, however, the simulation over-predicts the lensing signal in bin 3 significantly for $R<R_v$. Combining clusters in all four redshift bins, we find that the simulation also shows a 30\% excess compared to data, because the sample is dominated by the highest redshift bin. 
Due to the slightly more extended $R_v$ distribution in the real data compared to the mock, especially in the highest redshift bin (see Fig.~\ref{fig:void radius}), we check whether or not including a weight, based on the ratio of the two $R_v$ distributions, can reduce the difference between the data and mock. 
However, the inclusion of this weight does not change the signal significantly.

We characterise the consistency between simulation result and data using the lensing amplitude $A_{\kappa}$, where $A_{\kappa}=\kappa_{\rm data}/\kappa_{\rm th}$. Assuming Gaussian likelihoods with $\mathcal{L}\propto\exp(-\chi^2/2)$ and using the Jackknife covariance for the combined case, we find $A_{\kappa}=0.937\pm0.087$ for all voids and $A_{\kappa}=0.712\pm0.076$ for all clusters. Assuming independence, this difference is formally $1.9\sigma$, so hardly compelling evidence of an inconsistency; the combined result gives $A_{\kappa}=0.811\pm0.057$.

In H21, the measured angular cross-correlation between CMB lensing and galaxy overdensity also has a lower amplitude, $A_{\kappa}=0.901\pm0.026$, given the best-fit {\em Planck} 2018 cosmological parameters, $\sigma_8=0.811$ and $\Omega_m=0.315$. 
We further measure the angular cross-correlation $C_{\ell}^{g\kappa}$ of the mock and compare it with data. In order to account for the difference in galaxy bias, we include galaxy auto-correlation $C_{\ell}^{gg}$, and compare the bias-independent quantity \smash{$R=C_{\ell}^{g\kappa}/(C_{\ell}^{gg})^{-1/2}$}. The lensing amplitude $A_{\kappa}$ is then given by $A_{\kappa}=R^{\rm data}/R^{\rm mock}$. We compare the binned modes with $10\leq\ell<500$, assuming a diagonal covariance where the diagonal terms, following equations (12) and (13) in H21.
We obtain the following values for $A_{\kappa}$ in the four bins: $0.84\pm0.06$, $0.81\pm0.05$, $0.86\pm0.04$, $0.79\pm0.04$, and for the unbinned case, $A_\kappa=0.85\pm0.03$, consistent with the stacked result. This may suggest that the lower lensing signal is likely contributed by high density peaks.

In summary, the selection of superstructures based on the projected galaxy number density seems to pick up genuine superstructures. This is evident from the perfect agreement of galaxy number density profiles between our mocks and the observed data, and the reasonable agreement between the predicted CMB lensing convergence profiles and the observed versions. The latter suggests that the spatial variation of gravitational potentials are genuinely correlated with the superstructures. This gives us confidence in using them to study their expected temporal variations, which should be found via their ISW imprints on the CMB temperature maps. We therefore keep our catalogues fixed and `unblind' our analysis from this stage.

%lensing stacking here

\subsection{Stacked ISW map}

\begin{figure*}
\centering
 \includegraphics[width=\textwidth]{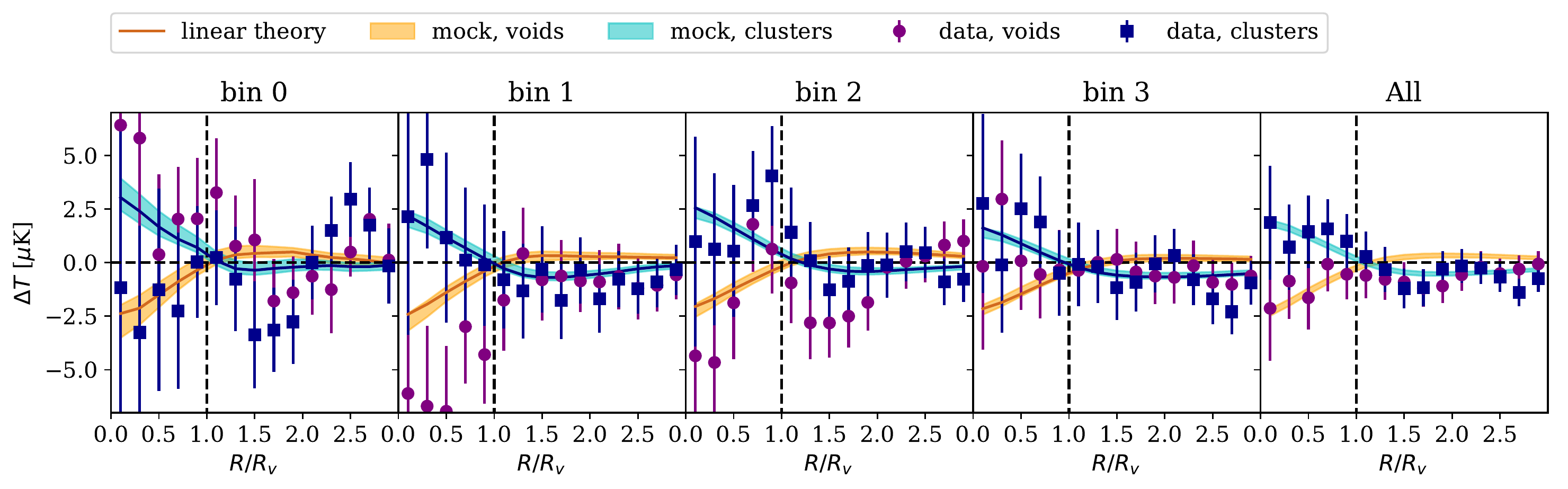}
 \caption{The averaged radial profile of stacked ISW temperature in each redshift bin at the superstructure positions found in the mock (solid band) and data (points). The quasi-linear predictions are shown by the thin lines. The {\em Planck} 2018 CMB temperature map is used for data with $\ell<10$ modes removed. The error bars come from Jackknife sampling of the superstructures in each redshift bin.}
 \label{fig:stacked profile isw}
\end{figure*}

We remove $\ell<10$ modes from the {\em Planck} 2018 CMB temperature map and the simulated ISW map to reduce the effect of the imperfectly simulated large-scale modes in the simulated ISW map as shown in Fig.~\ref{fig:ISW_map_autoCL_lin_nonlin_compare}.
A comparison of the stacked ISW profiles in data and simulation is presented in Fig.~\ref{fig:stacked profile isw}.
The quasi-linear theory prediction from the lensing potential gives consistent results, as does the full non-linear calculation, shown in the thin and broad solid lines respectively. 
Given the size of the error, the data shows general consistency with the simulation.
In the void case, it is noticeable that in bin 0, the measurement from data is slightly positive, whereas in bin 1, the data has a larger amplitude compared to the mock with $R<R_v$. The level of fluctuations in the four measurements suggest that these deviations are not statistically significant. We use the covariance matrix obtained from 1000 sets of random void positions to quantify the consistency between data and simulation. Given 15 degrees of freedom, the $\chi^2$ for each redshift bin is 8.9, 11.1, 16.2, and 11.8. The null test of the data signal gives $\chi^2$ of 8.1, 12.7, 15.2, and 10.2. In general, the data does not show a preference for the simulation prediction over a null signal.
For clusters, similar level of statistical fluctuations are present, with $\chi^2=11.3, 7.6, 10.8, 16.1$ for data compared to simulation, and $\chi^2=10.5,8.9,11.3,17.3$ for the null test.
Combining voids in all four bins, we find that $\chi^2=12.6$ for simulation and $\chi^2=10.1$ for a null signal. The larger $\chi^2$ for the simulation is probably due to the slightly negative signal at $R>R_v$. The combined cluster result shows $\chi^2=11.1$ for simulation and $\chi^2=15.1$ for null signal.
We characterise the consistency between simulation result and data using the ISW amplitude $A_{\rm ISW}$, where $A_{\rm ISW}=\Delta T_{\rm data}/\Delta T_{\rm th}$. Assuming Gaussian likelihoods with $\mathcal{L}\propto\exp(-\chi^2/2)$, we find $A_{\rm ISW}=-0.10\pm0.69$ for all voids and $A_{\rm ISW}=1.52\pm0.72$ for all clusters. The combined result gives $A_{\rm ISW}=0.68\pm0.50$.
Therefore, given the size of error, the measurements are fully consistent with the $\Lambda$CDM prediction; however, there is also no clear detection of this signal.

\subsection{Comparison with K19}\label{sec: Comparison with K19}

%our fiducial setting with Rv>100
We investigate the possible causes of the excess signal in K19. We attempt to apply the same void finding algorithm as K19, but with a few differences. Firstly, they used redshift bins with a comoving width of $100\mpcoh$ between $0.2<z<0.9$, whereas our bins are much wider in our fiducial setting. Secondly, due to the larger galaxy bias of the redMaGiC sample, they use a fixed $\delta_*=-0.3$ in the void finding algorithm across all redshift bins, and a comoving smoothing scale of $50\mpcoh$. In our fiducial setting, we have chosen to define $\delta_*$ to correspond to the lowest $10\%$ in density, and applied a comoving smoothing scale of $20\mpcoh$. Thirdly, in K19 a subsample of supervoids, with $R_v>100\mpcoh$ in particular, gave the excess signal, whereas in our fiducial void sample, we have not made selections based on void properties. Finally, the void sample in K19 is only within the DES footprint, whereas our sample covers a larger region.

To begin with, we make the assumption that differences in the void finding process would not lead to an inconsistent stacking signal, because the underlying structures found should correspond to the same physical underdensities. In this case, one possibility could be that the excess is only contributed by the supervoids with $R_v>100\mpcoh$. 

\begin{figure}
\centering
\includegraphics[width=0.47\textwidth]{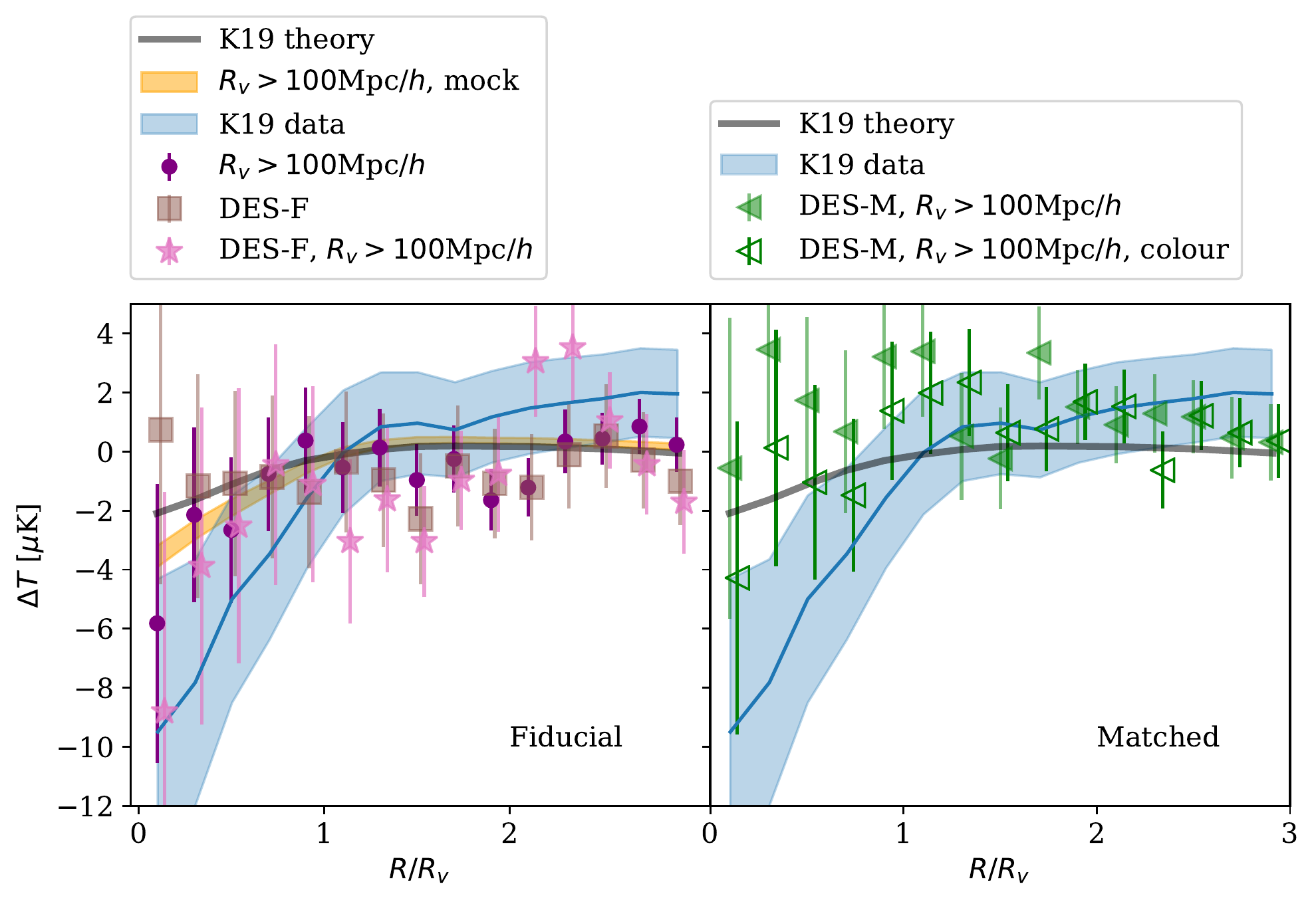}
 \caption{Stacked void profiles for a few subsamples chosen to match the K19 measurements (data shown as a blue band and theory shown as a grey solid line). The subsamples involving our fiducial setting are shown on the left panel, including: selection of void radius $R_v>100\mpcoh$ (purple circle); selection within the DES footprint (brown square); and selection within DES footprint as well as cut on $R_v$ (pink star). We also consider subsamples that are more closely matched to K19 in the void finding process within the DES footprint with and without a redMaGiC-like colour selection (shown on the right panel as open and filled green triangles). The error bars are given by Jackknife resampling.}
 \label{fig:stacked_profile_isw-scaled-DESfootprint-data-K19}
\end{figure}

Thus, we look at such subsample with our fiducial setting. 
This gives a total of 151 simulated voids, and 187 voids in the actual data. This number is smaller than one would expect from the K19 sample, which comprises 87 voids with $R_v>100\mpcoh$ within the DES footprint, if it were extended about 3 times to the same size as the Legacy Survey. This difference can be attributed to the thicker redshift slices used in our analysis. An additional factor is that most of the DES Y1 region is masked owing to our completeness cut, thus we may also lose a number of voids from that area. The stacked ISW profiles are shown on the right panel of Fig.~\ref{fig:stacked_profile_isw-scaled-DESfootprint-data-K19}. 
The overall signal from data (purple dots) shows good consistency with our simulation results (yellow band). On the same plot, we also copy the results from K19. While their theoretical prediction (grey solid line) seems to be smaller than ours, their void signal from the DES sample (blue band) is much stronger. The difference in the theoretical prediction is plausibly due to the difference involved in the void finding procedure. Using the covariance matrix from 1000 sets of random void samples, the $\chi^2$ is 16.3 compared to simulation and 16.5 compared to a null signal with ${\rm DOF=15}$. This suggests that in our fiducial sample, the large voids with $R_v>100\mpcoh$ do not cause an excess ISW signal.

%our fiducial setting inside DES footprint (test cosmic variance)
Another possibility is that the K19 excess is due to cosmic variance. To test this, we apply the same survey mask from the DES footprint, giving a subsample of 173 voids, with 40 voids among them having $R_v>100\mpcoh$. 
%this number is smaller than the K19 sample, which comprises 87 voids with $R_v>100\mpcoh$. %This difference can be attributed to the thicker redshift slices used in our analysis. An additional factor is that most of the DES Y1 region is masked owing to our completeness cut, and thus we may also lose a number of voids from that area. 
As shown on the right panel of Fig.~\ref{fig:stacked_profile_isw-scaled-DESfootprint-data-K19}, the stacked signal using all voids within the DES footprint (brown squares) is consistent with zero, but selecting the large voids (pink stars) does result in a mean signal closer to that measured in K19. However, given the size of the error bars, the overall signal is consistent with both a null signal and the simulation prediction.

%effect of bin size (show in the plot)
%We also tested the effect of the bin size. For example, we half the current bin size and repeat the above measurements. The result does not show clear signal in the DES footprint.

%\begin{figure*}
%\centering
% \includegraphics[width=0.8\textwidth]{plots/stacked_profile_isw-scaled-Rv100-exc200-data-mock-K19.pdf}
% \caption{{\em Left}: the 2D stacked CMB temperature map for voids in the DESI Legacy Survey with $R_v\geq100\mpcoh$. For display purpose only, each CMB map has been smoothed by a Gaussian kernel of $\sigma=2^{\circ}$. The inner and outer circles mark $R=R_v$ and $R=3R_v$. {\em Right}: the averaged radial profile of stacked ISW signal for voids with $R_v\geq100\mpcoh$ in simulation (yellow band) and data (purple points). In addition, we adopt the theory and measurements in \protect\cite{kovacs2019} here shown in blue solid line and blue triangle data points.}
% \label{fig:stacked profile isw Rv100}
%\end{figure*}

%Matched setting and mask: DESY1, DES full, and DESI Legacy Survey full
The above investigation suggests that the excess signal may be due to differences in the redshift binning and parameter choices in the void finding process. Thus, we try to follow the procedure outlined in K19 (and references therein) as closely as possible in order to see if we  can reproduce their signal.
We split our photometric sample in the redshift range $0.2<z<0.8$ into bins of comoving width of $100\mpcoh$. We exclude bins beyond $z\approx0.7$ due to a sharp drop in number density. This gives a total of 11 redshift bins. We also create another sample that has a matched colour distribution in $g-W_1$ vs $r-z$ and $g-r$ vs $r-z$ as the DESY1A1 RedMaGiC sample. The details of the selection criteria can be found in Appendix~\ref{apdx: Matching redMaGiC colour selection}. Such a selection removes rouhgly half of the sample compared to the unmatched one. 
To account for the masked DES Y1 region, we relax the completeness threshold for the mask to $30\%$ so that most of the DES Y1 region is now included. The completeness weighting and stellar density correction is then applied to each density map in a similar way to H21. Finally, due to the large photo-$z$ tail, we expect neighbouring bins to overlap significantly. In K19, a careful pruning of voids was applied by shifting the redshift binning by a small amount. In this case we apply a simplified version, where for neighbouring bins we remove the voids in the higher redshift bin if their centre lies within $0.5R_v$ of the voids in the lower redshift bin. We check that this removes most of the overlapping voids.
We also apply the same smoothing scale as in K19, $\sigma=50\mpcoh$, in void finding. 
We find 75 and 64 voids with $R_v>100\mpcoh$ inside the DES footprint with and without colour space constraints respectively, comparable to the 87 samples in K19. The stacked signal from these `matched' samples are shown on the left panel of Fig.~\ref{fig:stacked_profile_isw-scaled-DESfootprint-data-K19} as green open triangles (with colour selection) and filled triangles (without colour selection). These signals are slightly positive at $R<R_v$, and do not reproduce the excess signal shown in K19 (blue band). 
Thus, the excess signal may be due to other details in the void catalogue construction. For example, the small redshift bins can be affected by the uncertainty of our photo-$z$ sample, which has a median of $|\Delta z|=0.027$ but with a large non-Gaussian tail.

To summarize, we have attempted to compare the ISW signal from our void sample with K19, by investigating cuts on the void size, cosmic variance, and void-finding procedure. In the first two cases, we do not see a clear deviation from our simulation prediction based on the $\Lambda$CDM cosmology. In the last case, we obtain a signal that is consistent with $\Lambda$CDM, rather than roughly three times larger than the theoretical prediction from K19. This difference may be caused by details in the galaxy catalogue such as the galaxy sample and the photometric redshifts.

%This is still lower than the 87 sample in K19. A closer examination shows that in the DESY1 region, our completeness is low and most of the area is masked. Since we only include voids which overlaps with the mask for more than 70\%, we barely find any voids inside the DESY1 region. Another reason for possible difference is that the normalization for the density field is done for the whole Legacy Survey footprint, which may differ from that done only for the DES footprint. This changes the threshold for the void finding procedure. While the deepest voids are expected to be the same, the shallower voids may be missing. The two samples give consistent results which are also consistent with zero, even slightly positive at the centre.

\subsection{Searching for higher ISW signal}

\begin{figure}
\centering
 \includegraphics[width=0.47\textwidth]{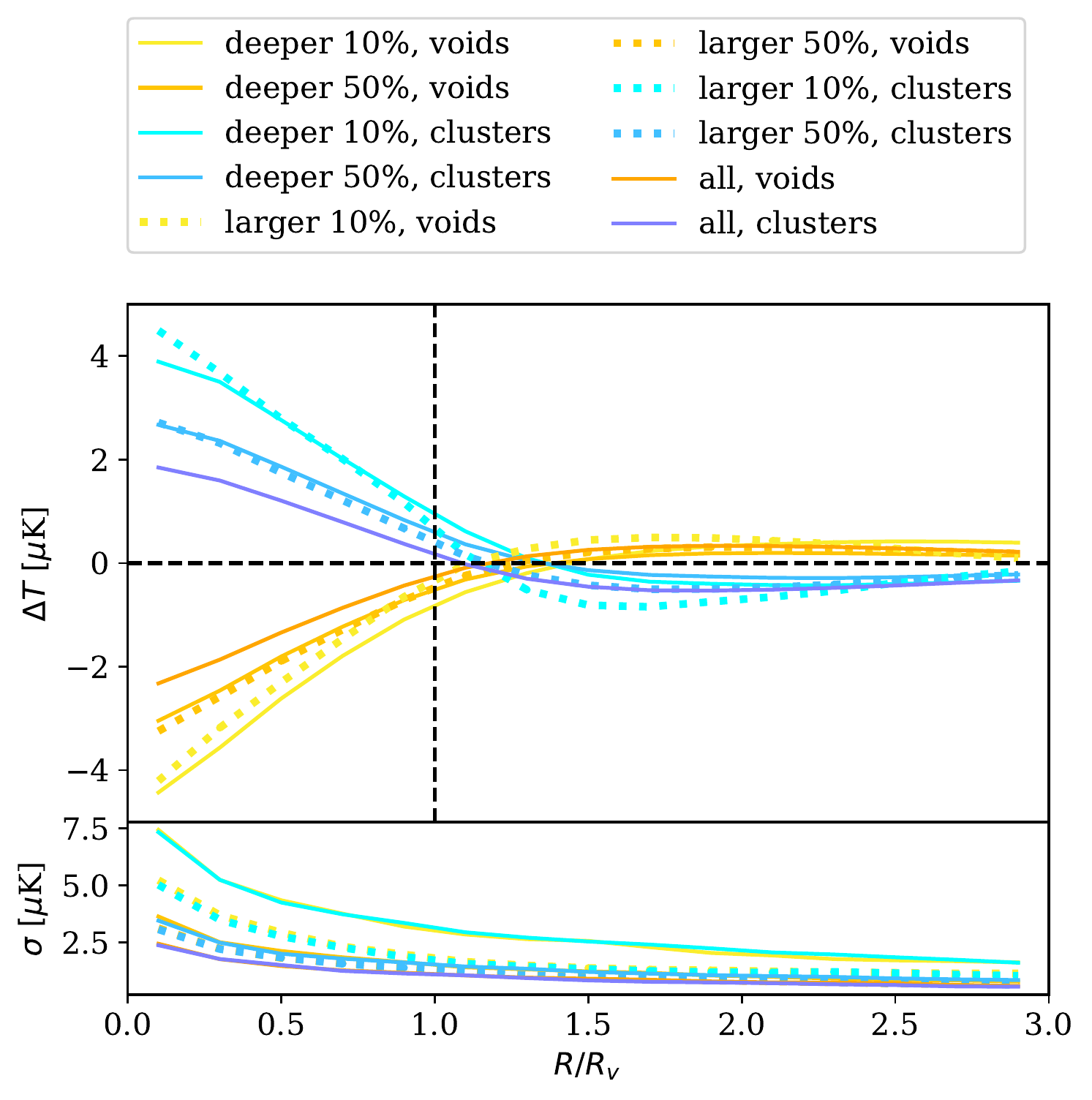}
 \caption{Stacked ISW profile split by central depth $\delta_c$ (solid line) and size $R_v$ (dotted line) using superstructures in the mock catalogue. The split is using the most extreme 10\%, 50\%, and the full sample in each case. Error bars are given by 1000 sets of random void stacking using the {\em Planck} CMB map.}
 \label{fig:stacked_profile_isw_mock_vary_dp_rv_quartile}
\end{figure}

In this section we look at the dependence of the signal-to-noise of the stacked ISW profile on supercluster properties. The purpose here is to see whether the excess ISW can be reproduced in by applying specific selections, rather than trying to claim a higher significance detection. Specifically we focus on $R_v$ and $\delta_c$, and in each case, we split the sample into the most extreme $10\%$ and $50\%$, and compare the SNR with the full sample. 
We use the simulation to determine the mean expected signal (thus the signal itself is noise free) and we show realistic errors by computing the covariance from 1000 sets of random void positions within the DESI Legacy Survey footprint, and stack using the {\em Planck} CMB map.
As shown in the upper panel in Fig.~\ref{fig:stacked_profile_isw_mock_vary_dp_rv_quartile}, selecting the 10\% most extreme objects in terms of $R_v$ or $\delta_c$ can boost the predicted ISW signal by about a factor of 2. From the lower panel in Fig.~\ref{fig:stacked_profile_isw_mock_vary_dp_rv_quartile}, it is clear that the larger $R_v$ has a smaller uncertainty compared to the more extreme $\delta_c$ selections with the same number of objects. This may be due to the fact that with the larger $R_v$ selection, the stacked profile is effectively averaging over a larger scale on the CMB map, thus reducing the noise on the profile. 

We measure the constraints on $A_{\rm ISW}$ for these selections in data. Focusing on the 10\% and 50\% of the superstructures with the largest $R_v$, we find that the data measurements show an increased signal especially in density peaks, with $A_{\rm ISW}=0.10\pm0.99, 0.57\pm0.71$ for voids and $A_{\rm ISW}=1.47\pm0.77, 2.59\pm0.73$ for clusters. Limiting the sample to the 10\% and 50\% with the most extreme $\delta_c$, we find that the data does not show a significant boost in the ISW signal, and $A_{\rm ISW}=0.15\pm1.24,0.32\pm0.89$ for voids and $A_{\rm ISW}=0.83\pm1.26,0.25\pm0.89$ for clusters.
Combining the voids and clusters in the $\delta_c$ selection, one finds $A_{\rm ISW}=0.75\pm0.83,0.58\pm0.59$ for the 10\% and 50\% of the total sample, which does not improve the significance of the signal compared to the full sample. On the other hand, in the $R_v$ case combining voids and clusters, $A_{\rm ISW}=0.96\pm0.61, 1.55\pm0.51$. 
The constraints on $R_v$ from the higher $R_v$ subsamples and the full sample is  statistically consistent with $0.3\sigma$ and $1.2\sigma$ for the 10\% and 50\% cases respectively. 
Therefore, by constraining on a larger $R_v$ sample, it is statistically possible to obtain a larger mean ISW signal, leading to a more significant detection of the ISW amplitude, $A_{\rm ISW}$. However, we emphasise that this selection is {\em a posteriori}, and one should take these results with caution.

%\textcolor{red}{Update: For the full sample, the ISW signal has $\chi^2=2.10$ from voids and $\chi^2=2.18$ from clusters with respect to a null detection with ${\rm DOF}=15$. Focusing on the 10\% and 50\% of the superstructures with largest $R_v$, we find $\chi^2=1.03, 1.97$ for voids and $\chi^2=1.90,2.08$ for clusters. Limiting the sample to the 10\% and 50\% with the most extreme $\delta_c$, we find $\chi^2=0.59,1.55$ for voids and $\chi^2=0.99,1.74$ for clusters. This test shows that to get a tighter constraint on $A_{\rm ISW}$, it is advantageous to increase the sample size than to condition on more extreme superstructures.} 

%\YC{(I suggest we do not include this section.)}

%Fig.~\ref{fig:stacked_profile_isw_mock_vary_dp_rv_quartile} shows the stacked ISW signal using the mock catalogue by splitting the void sample into quartiles ranked by void depth and void radius.
%In the split depth case, clearly there is a larger ISW signal coming from deeper voids. Notice that the ISW signal from deeper voids also tends to cross zero at $R>R_v$. In the split radius case, the trend is not so clear. From small to large void radius, the signal amplifies and the zero crossing scale increases. However, in the largest radius bin, this trend is slightly reversed. By examining the stacked density profiles of these largest voids, we find that they are likely to overlap with deeper voids found within their radius.

\section{Conclusions}
\label{sec: conclusions}

In this work we have constructed a catalogue of superstructures, using tomographic data with $0<z<0.8$ in the DESI Legacy Imaging Survey. We adopt the void finding algorithm described in \cite{sanchez2017}, taking the lowest 10\% and highest 5\% pixels of the galaxy density field after 2D Gaussian smoothing with $\sigma=20\mpcoh$. The aim has been to test the excess ISW signal from supervoids claimed in literature \citep{2008ApJ...683L..99G, kovacs2019}. To compare our results with the $\Lambda$CDM model prediction, we constructed a mock catalogue using the {\em Multidark} simulation. The galaxy number density, linear, and non-linear galaxy biases are matched to those found in our previous work on the DESI Legacy Imaging Survey (\citejap{hang2021}: H21), and we applied a redshift error to match the photo-$z$ precision found in H21. The properties of the superstructures and the stacked galaxy density profiles around these superstructures are consistent between the mock and the data. We also created the corresponding lensing convergence and ISW maps. 

Subsequently, we looked at the stacked CMB lensing convergence and CMB temperature using the {\em Planck} 2018 maps at the centre of these superstructures, scaled by the void/cluster radius $R_v$. The comparison between the stacked lensing signal agrees well in the void case, but the cluster signal seems to be slightly over-predicted in the highest redshift bin. Using the covariance matrix from 1000 sets of randomized superstructure positions, we quantify the consistency between simulation and data via the lensing amplitude $A_\kappa$, and find $A_\kappa=0.81\pm0.06$ from combining the voids and clusters. This is largely driven by the highest redshift bin, which contains the most clusters. In H21, we favoured a lensing amplitude of $A_\kappa=0.93\pm0.03$ compared to the theoretical prediction from the {\em Planck} 2018 best-fit cosmology, using cross-correlation in spherical harmonic space (therefore essentially utilizing all pixels, rather than density peaks and troughs). The amplitude of the CMB lensing signal is consistent with our result from superstructures, although we note that the voids and clusters are in $1.9\sigma$ tension, with $A_{\kappa}=0.937\pm0.087$ for all voids and $A_{\kappa}=0.712\pm0.076$ for all clusters. Despite this, the level of disagreement between our mocks and data for the lensing signal is negligibly small for the purpose of the ISW study, as its measurement is much noisier.

The stacked ISW signals are in general consistent with the simulation results -- but also with a null signal, reflecting the low signal-to-noise of the ISW effect. Specifically, we do not detect a significant signal from the void catalogue, and only a marginal signal from clusters. Combining the superstructures, we find the ISW amplitude to be $A_{\rm ISW}=0.68\pm0.50$, somewhat weaker than the cross-correlation result from H21 which gave $A_{\rm ISW}=1.10\pm0.31$ (although both measurements are consistent). Therefore, we do not claim a detected ISW signal using this sample.

We compare our results with K19, \cite{kovacs2019}, who reported a $3\sigma$ excess ISW signal compared to the $\Lambda$CDM prediction from supervoids with void radius $R_v>100\mpcoh$, using the DES redMaGiC sample within a similar redshift range to ours. 
Using our fiducial settings described above, we do not find any excess signal from voids with the same size cut, or within the same survey mask.
Two subsamples are then constructed to match the redshift binning and void finding procedure in K19 as closely as possible within the DES footprint, with and without a redMaGiC-like colour selection. The stacked ISW profiles from the voids found in these samples also do not show any anomalous signal.

Lastly, we look at the dependence of the ISW signal on the void properties and discuss whether this can be used to explain a higher detection of the ISW signal from suitably chosen superstructures. We show that the mean ISW signal from the mock dataset is amplified by excluding smaller or less extreme superstructures, while the shot noise increases. Applying the same selections to the data, we find no significant improvement in the constraint on the ISW amplitude $A_{\rm ISW}$ from more extreme superstructures, although there is a boost in $A_{\rm ISW}$ from density peaks with larger $R_v$. The most extreme subset conditioning on the 10\% largest $R_v$ gives $A_{\rm ISW}=0.96\pm0.61$, and the 95\% upper limit is 2.16. However, we emphasise that the selection of this subset is not {\em a priori}. The increase in the mean signal is therefore inevitably overestimated due the look-elsewhere effect. 
%The measurements of $A_{\rm ISW}$ from different subsamples are not in statistical tension, indicating that it is statistically possible to see a higher than usual ISW signal from specific selections.

In summary, then, our results from investigating the impact of superstructures on the CMB do not reveal any compelling discrepancy with $\Lambda$CDM. The CMB lensing results for superstructures independently favour an approximately 10\% reduction in amplitude relative to the {\em Planck} 2018 prediction, in very close agreement with our conclusion from the overall galaxy-lensing cross-correlation in H21, which we argued favoured a matter density at the low end of the range permitted by {\em Planck}. The evidence for this reduced lensing amplitude is present in both voids and clusters although the latter favour a stronger signal at the $1.9\sigma$ level; it will be interesting to see if this tension becomes more significant in future datasets.
Similarly, the ISW signal from stacked superstructures is consistent with the H21 cross-correlation result, and not in significant disagreement with $\Lambda$CDM. Formally, the 95\% confidence upper limit on $A_{\rm ISW}$ from superstructures is 1.51, and therefore we do not reproduce literature claims of anomalous superstructure ISW signals at several times the $\Lambda$CDM prediction. We have tried to vary our analysis in order to mimic more closely the selection involved in these claims, but have not succeeded in raising the ISW signal. Presumably some small differences in method remain. But the important point is that any such excess is apparently not robust, since we were not able to produce an excess signal even by exploring a number of alternative forms of superstructure selection. 

A similar investigation has been performed by \cite{2021MNRAS.500.3838D}, who measured the cross-correlation between the `low-density position' (LDP) and the CMB using the DESI Legacy Survey. In this work, they detected the ISW effect from their underdense regions with a significance of $3.4\sigma$, and this signal is fully consistent with the $\Lambda$CDM prediction.

\section*{Acknowledgements}

QH was supported by the Edinburgh Global Research Scholarship and the Higgs Scholarship from Edinburgh University. SA and JAP were supported by the European Research Council under grant number 670193 (the COSFORM project). YC acknowledges the support of the Royal Society through the award of a University Research Fellowship and an Enhancement Award.

The Legacy Surveys consist of three individual and complementary projects: the Dark Energy Camera Legacy Survey (DECaLS; NOAO Proposal ID \# 2014B-0404; PIs: David Schlegel and Arjun Dey), the Beijing-Arizona Sky Survey (BASS; NOAO Proposal ID \# 2015A-0801; PIs: Zhou Xu and Xiaohui Fan), and the Mayall z-band Legacy Survey (MzLS; NOAO Proposal ID \# 2016A-0453; PI: Arjun Dey). DECaLS, BASS and MzLS together include data obtained, respectively, at the Blanco telescope, Cerro Tololo Inter-American Observatory, National Optical Astronomy Observatory (NOAO); the Bok telescope, Steward Observatory, University of Arizona; and the Mayall telescope, Kitt Peak National Observatory, NOAO. The Legacy Surveys project is honored to be permitted to conduct astronomical research on Iolkam Du'ag (Kitt Peak), a mountain with particular significance to the Tohono O'odham Nation.

NOAO is operated by the Association of Universities for Research in Astronomy (AURA) under a cooperative agreement with the National Science Foundation.

This project used data obtained with the Dark Energy Camera (DECam), which was constructed by the Dark Energy Survey (DES) collaboration. Funding for the DES Projects has been provided by the U.S. Department of Energy, the U.S. National Science Foundation, the Ministry of Science and Education of Spain, the Science and Technology Facilities Council of the United Kingdom, the Higher Education Funding Council for England, the National Center for Supercomputing Applications at the University of Illinois at Urbana-Champaign, the Kavli Institute of Cosmological Physics at the University of Chicago, Center for Cosmology and Astro-Particle Physics at the Ohio State University, the Mitchell Institute for Fundamental Physics and Astronomy at Texas A\&M University, Financiadora de Estudos e Projetos, Fundacao Carlos Chagas Filho de Amparo, Financiadora de Estudos e Projetos, Fundacao Carlos Chagas Filho de Amparo a Pesquisa do Estado do Rio de Janeiro, Conselho Nacional de Desenvolvimento Cientifico e Tecnologico and the Ministerio da Ciencia, Tecnologia e Inovacao, the Deutsche Forschungsgemeinschaft and the Collaborating Institutions in the Dark Energy Survey. The Collaborating Institutions are Argonne National Laboratory, the University of California at Santa Cruz, the University of Cambridge, Centro de Investigaciones Energeticas, Medioambientales y Tecnologicas-Madrid, the University of Chicago, University College London, the DES-Brazil Consortium, the University of Edinburgh, the Eidgenossische Technische Hochschule (ETH) Zurich, Fermi National Accelerator Laboratory, the University of Illinois at Urbana-Champaign, the Institut de Ciencies de l'Espai (IEEC/CSIC), the Institut de Fisica d'Altes Energies, Lawrence Berkeley National Laboratory, the Ludwig-Maximilians Universitat Munchen and the associated Excellence Cluster Universe, the University of Michigan, the National Optical Astronomy Observatory, the University of Nottingham, the Ohio State University, the University of Pennsylvania, the University of Portsmouth, SLAC National Accelerator Laboratory, Stanford University, the University of Sussex, and Texas A\&M University.

BASS is a key project of the Telescope Access Program (TAP), which has been funded by the National Astronomical Observatories of China, the Chinese Academy of Sciences (the Strategic Priority Research Program "The Emergence of Cosmological Structures" Grant \# XDB09000000), and the Special Fund for Astronomy from the Ministry of Finance. The BASS is also supported by the External Cooperation Program of Chinese Academy of Sciences (Grant \# 114A11KYSB20160057), and Chinese National Natural Science Foundation (Grant \# 11433005).

The Legacy Survey team makes use of data products from the Near-Earth Object Wide-field Infrared Survey Explorer (NEOWISE), which is a project of the Jet Propulsion Laboratory/California Institute of Technology. NEOWISE is funded by the National Aeronautics and Space Administration.

The Legacy Surveys imaging of the DESI footprint is supported by the Director, Office of Science, Office of High Energy Physics of the U.S. Department of Energy under Contract No. DE-AC02-05CH11231, by the National Energy Research Scientific Computing Center, a DOE Office of Science User Facility under the same contract; and by the U.S. National Science Foundation, Division of Astronomical Sciences under Contract No. AST-0950945 to NOAO.

The CosmoSim database used in this paper is a service by the Leibniz-Institute for Astrophysics Potsdam (AIP).
The MultiDark database was developed in cooperation with the Spanish MultiDark Consolider Project CSD2009-00064.

The authors gratefully acknowledge the Gauss Centre for Supercomputing e.V. (www.gauss-centre.eu) and the Partnership for Advanced Supercomputing in Europe (PRACE, www.prace-ri.eu) for funding the MultiDark simulation project by providing computing time on the GCS Supercomputer SuperMUC at Leibniz Supercomputing Centre (LRZ, www.lrz.de).

%%%%%%%%%%%%%%%%%%%%%%%%%%%%%%%%%%%%%%%%%%%%%%%%%%
\section*{Data Availability}

All of the observational datasets used in this paper are available through the Legacy Survey website \url{http://legacysurvey.org/dr8/}. The codes used in this analysis along with several processed data products can be accessed at \url{https://gitlab.com/qianjunhang/desi-legacy-survey-superstructure-stacking}.

%%%%%%%%%%%%%%%%%%%% REFERENCES %%%%%%%%%%%%%%%%%%

% The best way to enter references is to use BibTeX:

\bibliographystyle{mnras}
\bibliography{project} % if your bibtex file is called example.bib

% Alternatively you could enter them by hand, like this:
% This method is tedious and prone to error if you have lots of references
%\begin{thebibliography}{99}
%\bibitem[\protect\citeauthoryear{Author}{2012}]{Author2012}
%Author A.~N., 2013, Journal of Improbable Astronomy, 1, 1
%\bibitem[\protect\citeauthoryear{Others}{2013}]{Others2013}
%Others S., 2012, Journal of Interesting Stuff, 17, 198
%\end{thebibliography}

%%%%%%%%%%%%%%%%%%%%%%%%%%%%%%%%%%%%%%%%%%%%%%%%%%

%%%%%%%%%%%%%%%%% APPENDICES %%%%%%%%%%%%%%%%%%%%%

\appendix

\section{Matching redMaGiC colour selection}\label{apdx: Matching redMaGiC colour selection}

\begin{figure*}
\centering
\includegraphics[width=0.7\textwidth]{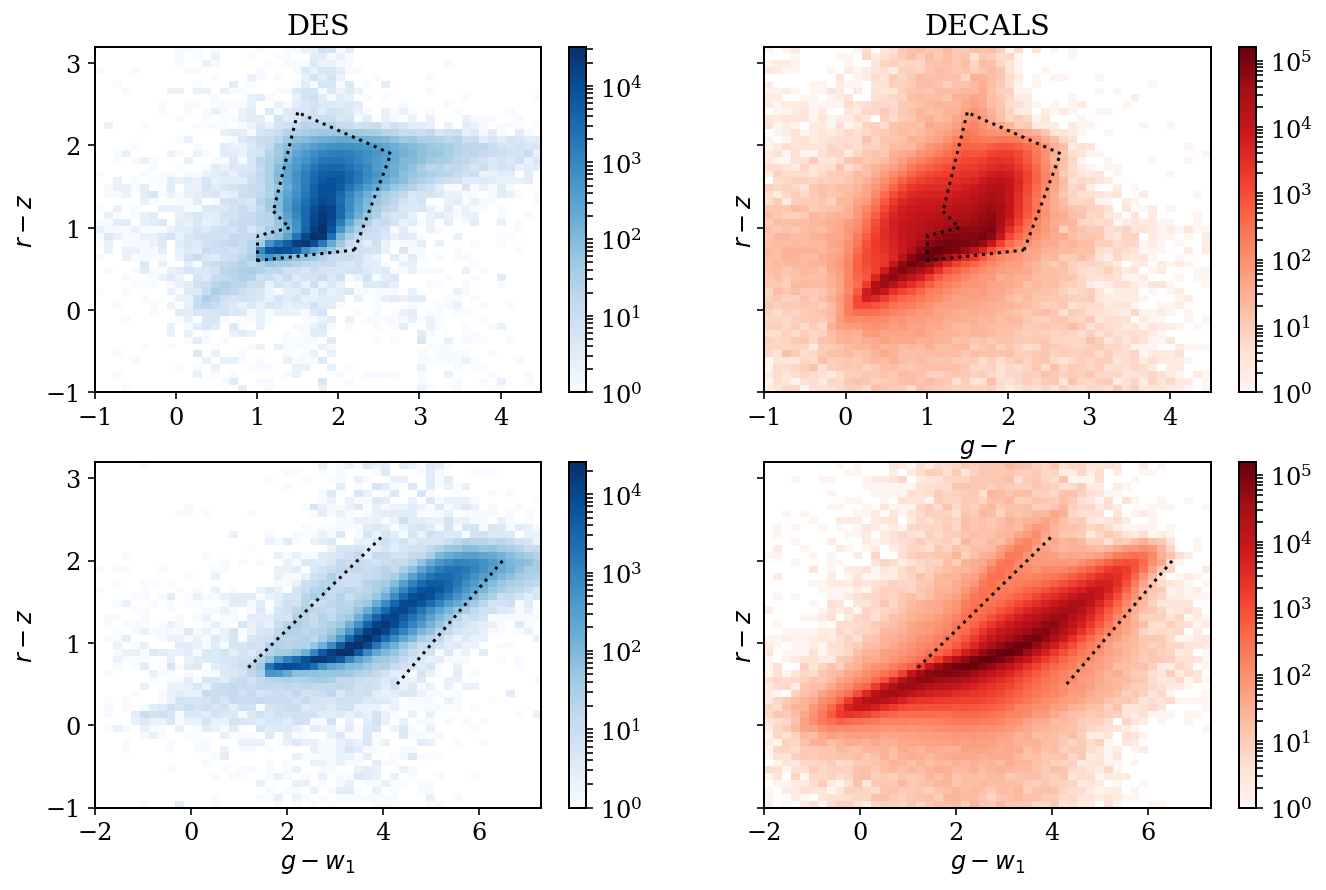}
\caption{The comparison of redMaGiC (left, blue) and DECALS (right, red) samples in the same sky area in $g-r$ and $r-z$ plane (upper panel), and in $g-w_1$ and $r-z$ plane (lower panel). DECALS contains a large number of bluer objects compared to redMaGiC. The thin strip on the left side of the main sequence is likely to be residual stars. The black dotted box is the region used to take ratios.}
\label{fig: colour_colour_plane_raw}
\end{figure*}

\begin{figure*}
\centering
\includegraphics[width=0.7\textwidth]{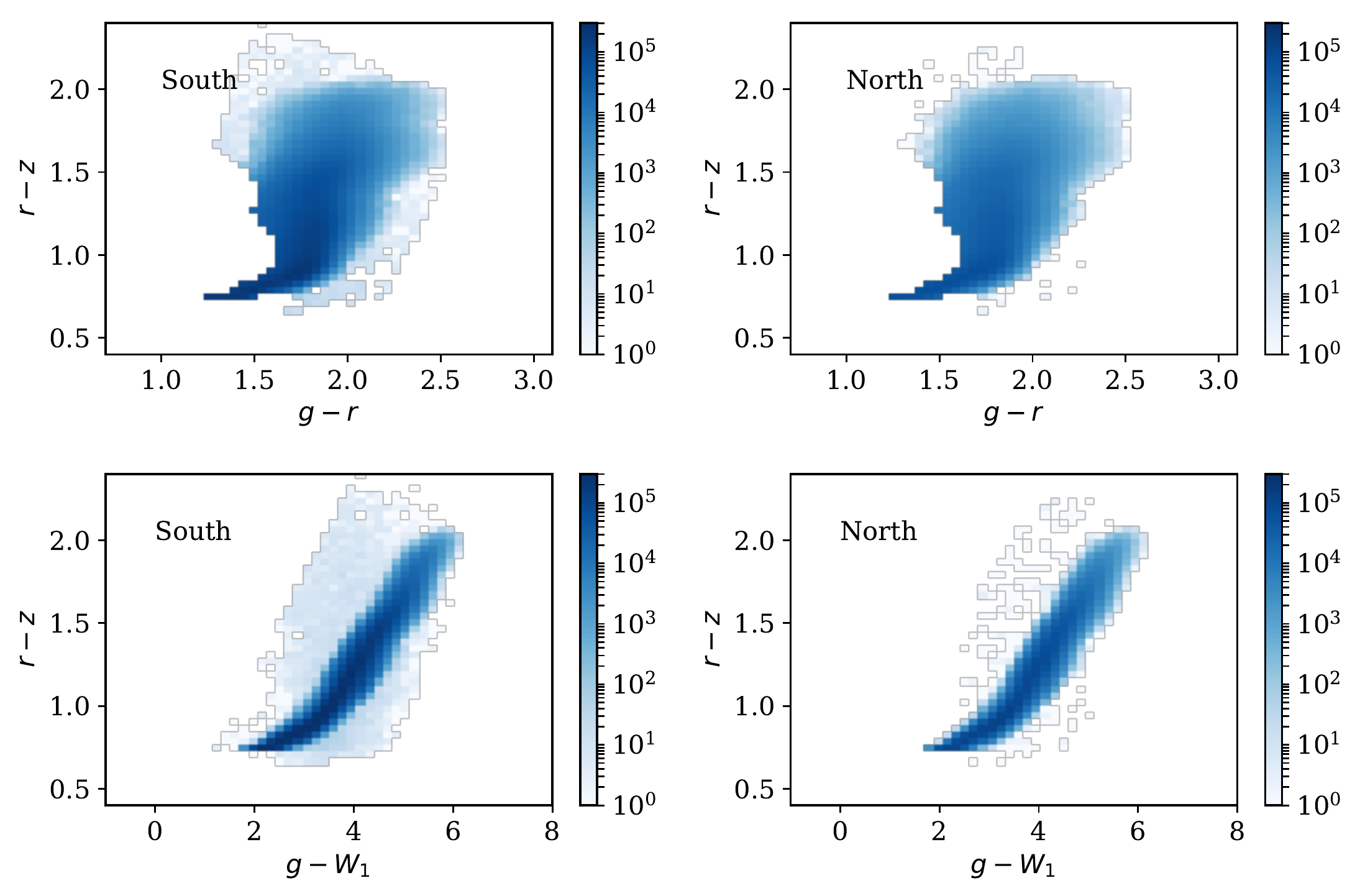}
\caption{The selection in $g-r$ vs $r-z$ and $g-W_1$ vs $r-z$ colour space for the DESI Legacy Survey galaxy sample in the north and south regions respectively to match the DESY1A1 redMaGiC sample.}
\label{fig: redmagic-sel-colour-plane}
\end{figure*}

In order to match the DESY1A1 redMaGiC galaxies as closely as possible, we compare their distribution in colour-colour space with a subsample of DECaLS galaxies in the same region (Fig.~\ref{fig: colour_colour_plane_raw}). We apply cuts in the $g-r$ versus $r-z$ plane based on the ratio of the normalized distribution. We exclude regions in this space where the ratio is smaller than a threshold set to 0.5. Such a exclusion does not affect the redMaGiC sample (about 92\% of our objects remain), but it results in a cut in low-redshift DECALS galaxies. The selected DECALS sample contains 1.8 million galaxies, about 3 times the redMaGiC sample. Fig.~\ref{fig: redmagic-sel-colour-plane} shows the selected region in the colour-colour plane for our full sample used in Section~\ref{sec: Comparison with K19} in the redshift range $0.2<z<0.8$ in the north and south part of the DESI Legacy Survey.

%%%%%%%%%%%%%%%%%%%%%%%%%%%%%%%%%%%%%%%%%%%%%%%%%%

% Don't change these lines
\bsp	% typesetting comment
\label{lastpage}
\end{document}